\documentclass{aa}  
\usepackage{lscape}
\usepackage{graphicx}
\usepackage{txfonts}
\usepackage{comment}
\usepackage{booktabs}
\usepackage{colortbl}
\usepackage{xcolor}
\usepackage[hidelinks,colorlinks=true,linkcolor=blue,citecolor=blue,urlcolor=black]{hyperref}
\usepackage{makecell}

\begin{document}

   \title{
   CHEX-MATE: Scaling relations of radio halo profiles for clusters in the LoTSS DR2 area 
   }
   \authorrunning{M. Balboni et al.}

   \author{M. Balboni
          \inst{\ref{iasf}, \ref{insubria},\ref{unibo}},
          S. Ettori
          \inst{\ref{oas},\ref{infn_bo}},
          F. Gastaldello \inst{\ref{iasf}},
          R. Cassano\inst{\ref{ira}},
          A. Bonafede\inst{\ref{ira},\ref{unibo}},
          V. Cuciti\inst{\ref{ira},\ref{unibo}},
          A. Botteon\inst{\ref{ira}},  
          G. Brunetti\inst{\ref{ira}},
          I. Bartalucci\inst{\ref{iasf}}, 
          M. Gaspari\inst{\ref{unimore}},
          R. Gavazzi\inst{\ref{marseille},\ref{sorbonne}}
          S. Ghizzardi\inst{\ref{iasf}}, 
          M. Gitti\inst{\ref{unibo},\ref{ira}}, 
          L. Lovisari\inst{\ref{iasf}}, 
          B. J. Maughan\inst{\ref{hhw}},
          S. Molendi\inst{\ref{iasf}}, 
          E. Pointecouteau\inst{\ref{irap}}, 
          G.W. Pratt\inst{\ref{saclay}},
          E. Rasia\inst{\ref{trieste}, \ref{michigan}},
          G. Riva\inst{\ref{iasf}, \ref{unimi}}, 
          M. Rossetti\inst{\ref{iasf}}, 
          H. Rottgering\inst{\ref{leiden}},
          J. Sayers\inst{\ref{caltech}},
          R.J. van Weeren\inst{\ref{leiden}}}

\institute{
        INAF - IASF Milano, via A. Corti 12, 20133 Milano, Italy \label{iasf}
        \and
        DiSAT, Universit\`a degli Studi dell’Insubria, via Valleggio 11, I-22100 Como, Italy \label{insubria}
        \and
        DIFA - Universit\`a di Bologna, via Gobetti 93/2, I-40129 Bologna, Italy\label{unibo}
        \and
        INAF, Osservatorio di Astrofisica e Scienza dello Spazio, via Piero Gobetti 93/3, 40129 Bologna, Italy\label{oas}
        \and
        INFN, Sezione di Bologna, viale Berti Pichat 6/2, 40127 Bologna, Italy\label{infn_bo}
        \and
        INAF - IRA, Via Gobetti 101, I-40129 Bologna, Italy\label{ira}
        \and
        Department of Physics, Informatics and Mathematics, University of Modena and Reggio Emilia, 41125 Modena, Italy\label{unimore}
        \and
        Laboratoire d’Astrophysique de Marseille, CNRS, Aix-Marseille Université, CNES, Marseille, France\label{marseille}
        \and
        Institut d’Astrophysique de Paris, UMR 7095, CNRS, and Sorbonne Université, 98bis boulevard Arago, 75014 Paris, France\label{sorbonne}
        \and
        HH Wills Physics Laboratory, University of Bristol, Tyndall, Bristol, BS8 1TL, UK\label{hhw}
        \and
        IRAP, Université de Toulouse, CNRS, CNES, UT3-UPS, (Toulouse), France\label{irap}
        \and
        Universit\'e Paris-Saclay, Universit\'e Paris Cit\'e, CEA, CNRS, AIM, 91191, Gif-sur-Yvette, France \label{saclay}
        \and
        INAF - Osservatorio Astronomico di Trieste, via Tiepolo 11, I-34143 Trieste, Italy\label{trieste}
        \and
        Department of Physics; University of Michigan, 450 Church St, Ann Arbor, MI 48109, USA\label{michigan}
        \and 
        Dipartimento di Fisica, Universitá degli Studi di Milano, Via G. Celoria 16, 20133 Milano, Italy \label{unimi}
        \and
        Leiden Observatory, Leiden University, PO Box 9513, 2300 RA Leiden, The Netherlands\label{leiden}
        \and
        California Institute of Technology, 1200 East California Boulevard, Pasadena, CA 91125, USA\label{caltech}
        }

   \date{}

 
    \abstract
    {The thermal and non-thermal components in galaxy clusters have properties that, although shaped from different physical phenomena, can share some similarities, mainly driven by their halo mass and the accretion processes. Scaling relations have been proven to exist for both components and studied in X-ray (thermal) and radio (non-thermal) bands. On the X-ray side, both integrated and spatially resolved profiles have shown a predictable and correlated behaviour. At the radio wavelength, such investigations are so far limited to the integrated quantities (e.g. total power and mass). We aimed to investigate the scaling relations between the mass of a galaxy cluster and its radio emission at low frequencies, treating both the integrated and the spatially resolved quantities for a sample of well-selected objects in a self-consistent analysis.
    We crossmatched LoTSS DR2 and CHEX-MATE datasets in order to get the deepest and most homogeneous radio data of a representative sample of objects.  Among the 40 CHEX-MATE objects in the LOFAR DR2 area, we investigated the 18 objects showing radio halo emission, which span a broad mass range, by extracting and analysing their radio emission profiles.
    We analytically derived the expected relation between the radio power ($P_{\nu}$) and radio surface brightness profile ($I_R(r)$),
    and performed a comparison with observational results.
    We obtained that properly accounting for the mass and redshift dependence in the radio profile can reduce the overall scatter by a factor of $\sim 4$, 
    with an evident residual dependence on the cluster dynamical status. 
    We show that assuming the halo size $R_{H} \sim R_{500}$ did not allow us to reconcile the expected (from our analytical derivations) and observed mass profile scaling.
    Instead, accounting for no $R_H - M$ relation, allowed us to reconcile the observed radio profile mass scaling and the one predicted starting from the $P_{\nu}-M$ relation.
    We discuss the implications of a lack of $R_H-M$ relation, assessing possible systematics and biases in the analyses, and interpreting it as a natural consequence of the structure formation process, where the halo size depends on both the cluster dynamical status, related to the strength of the merger, and mass. 
    Finally, we also considered the role of the magnetic field in the $P_{\nu}-M$ relation, putting constraints on its dependence upon the cluster mass and finding consistent results with expectations from our radio power mass scaling.}

   \keywords{         }

   \maketitle
%

\section{Introduction}\label{sec:intro}

Galaxy cluster formation occurs via the hierarchical growth of sub-structures and the diffuse medium driven by the gravitational field of the dark matter component. The baryonic matter that accretes onto these structures is heated up to $10^7 - 10^8$ K (depending on the halo mass) through shocks, forming a hot and rarefied ($n_{gas} \sim 10^{-4} - 10^{-2} ~ \rm{cm^{-3}}$) plasma, the intra-cluster medium (ICM). As a consequence of its temperature and density, the ICM emits via thermal Bremsstrahlung in the X-rays, making galaxy clusters appear as diffuse sources in the sky.\par
Radio observations have shown that several clusters also present a diffuse, non-thermal ICM component that emits synchrotron radiation. Such emission traces cosmic ray electrons (CRe) and weak magnetic fields ($\sim \mu$G) throughout the whole extension of the clusters. 
Depending on the particle acceleration mechanism, there is a variety of diffuse radio sources in clusters: giant radio halos, mini-halos, radio relics, and revived fossil plasma (see \citealt{vanWeeren19} for a review).
Giant radio halos are among the most extended cases of such diffuse emission.
They are roundish, megaparsec-scale radio sources, characterised by a steep spectral index ($\alpha>1$ where the flux $S_{\nu} \propto \nu^{-\alpha}$). Radio halos are mostly found in massive merging clusters and are co-spatial with the thermal plasma.
The commonly accepted scenario is that they are the result of turbulent re-acceleration, induced by merger events, of mildly relativistic electrons already present in the ICM due to past energetic phenomena, e.g. active galactic nuclei (AGN) activity or previous mergers (\citealt{Brunetti14}).\par 
It has been established early that the radio power of giant halos correlates with the X-ray luminosity of their host clusters \citep[e.g.][]{Liang2000,Cassano06,Kale2015} and that there is a bi-modality in the $P_{\nu}-L_X$ plane, with clusters without a radio halo, and more dynamically relaxed, populating a region well separated from the correlation \citep{Brunetti2009}, further strengthening the merger-radio halo connection.
Since the advent of Sunyaev-Zel'dovich (SZ, \cite{SZ1972}) surveys, the integrated SZ signal, which arises from the integrated pressure along the line of sight of the ICM, shows a similar tight correlation with the radio power \citep[e.g.][]{Basu2012,Cassano2013}. Both quantities are a proxy for the cluster mass, with the SZ signal being less scattered with respect to the X-ray luminosity, revealing the underlying relation between the radio power and mass.
The reason for that relation arises naturally in the context of the turbulent re-acceleration scenario since the energy available for re-acceleration is a fraction of the gravitational energy released during a merger event \citep{Cassano2005,Cassano2007,Eckert2017}.
With the advent of a new generation of radio telescopes and, in particular, the ones operating at a low frequency, the halo radio power-mass relation ($P_{\nu} - M$) has been established and extended from the 1.4 GHz frequency range of the early studies down to the 150 MHz frequency range 
\citep{vanWeeren2021,Cuciti2023}.\par   
The existence of a scaling law points to an underlying physical mechanism producing it: explaining its origin with a minimal set of theoretical assumptions holds the promise of gaining critical physical insight. We can try to look here with a hopefully enriching analogy to the current landscape of the investigations of scaling relations concerning the physical properties of the thermal component, the ICM. 
The relations between the observables (e.g. $L_X$ and $T_{ICM}$) and mass do not follow the expected self-similar scaling where gravity dominates \citep[e.g.][]{Voit05}, and can be further modified due to astrophysical processes (e.g. AGN feedback, \citealt{Gaspari2020}). 
The use of multi-variate scaling relations between different physical quantities and mass, and simple assumptions such as the role of the gas fraction and gas clumpiness, can provide a coherent picture of the modifications to the self-similar scenario needed to explain the observed behaviour of the scaling relations \citep[see e.g.][]{arnaud99,Pratt2010,Ettori2015}. As a final step, the observed radial profiles and the integrated values of the thermodynamic quantities can be linked, leading to self-consistent predictions for their behavior in various observational samples \citep{Ettori2020,Pratt2022,Ettori2023}.\par
In a similar fashion, for the first time for the non-thermal component in galaxy clusters, 
we attempted to link the radio surface brightness profiles to the scaling between the global radio power and the cluster mass. 
We performed the exercise on the galaxy clusters observed in the LOFAR Two-meter Sky Survey Data Release 2 \citep[LoTSS DR2,][]{Shimwell2022}, which are also part of the Cluster HEritage project with XMM-Newton - Mass Assembly and Thermodynamics at the Endpoint of structure formation \citep[CHEX-MATE,][]{Arnaud21}.  This is part of a project to fully characterise the non-thermal properties of the objects in the sample hosting radio halos \citep[see][for a pilot study]{Balboni2024}.\par
With the aim of helping the readers navigate through the text, we provide a detailed description of its structure in the following paragraphs. In Sect.~\ref{sec:sample_datared}, we describe the sample and briefly report the data reduction.
In Sect.~\ref{sec:analysis}, we present the data analysis following a step-by-step approach. 
We start by introducing a model that allows us to link the cluster's $P_{\nu}-M$ relation to the radio halo profile properties, obtaining a scaling in both mass and redshift (Sect.~\ref{sec:model}).
In Sect.~\ref{sec:m_only}, we derive model predictions and, in Sect.~\ref{sec:prof_scaling}, we test those using observational data. We start by fitting, with a simple version of the presented model, the observed $P_{\nu}-M$ relation and use such results to derive the expected mass dependence of the radio halo profiles.
We then compare the expected halo profile scaling with the one recovered through observations. 
We progressively increase the model complexity to find a good agreement between the predicted and observed relations among radially dependent and integrated quantities. 
In Sect.~\ref{sec:no_RH-M}, we relax some of the assumptions made on the scaling between halo quantities, such as the cluster mass and the size of the halo, to better reproduce the observational results. 
In Sect.~\ref{sec:discussion}, we present and discuss the main findings of our study.
In particular, in Sect.~\ref{sec:P-MB} we add the dependence of the cluster magnetic field to the model, gaining insight on the scaling of the magnetic field with the mass.
We summarise the work done and present future perspectives in Sect.~\ref{sec:conclusion}.\\
Throughout this work we assume a flat, $\Lambda$CDM Universe cosmology with $H_0 = 70 \rm ~ km/s/Mpc$, $\Omega_{m,0}=0.3$, $\Omega_{\Lambda} = 0.7$, $H_z / H_0 = E_z = [ \Omega_{m, 0} (1 + z)^3 + \Omega_{\Lambda} ]^{1/2}$.

\section{Sample overview and data reduction}\label{sec:sample_datared}

\subsection{Description of the samples}\label{sec:sample}

The cluster sample we use here derives from the combination of two datasets: CHEX-MATE and the LoTSS DR2. The CHEX-MATE project \citep{Arnaud21} is a three mega-second XMM-Newton Multi-Year Heritage Programme to obtain X-ray observations of a minimally-biased, signal-to-noise limited sample of 118 galaxy clusters detected by \textit{Planck} through the Sunyaev-Zel'dovich effect \citep{Planck2016}. 
The programme aims to study the ultimate products of structure formation in time and mass, using a census of the most recent objects to have formed (Tier-1: $0.05 < {z} < 0.2$ ; ${M_{500}}$\footnote{$M_{500} = 500 \frac{4}{3} \rho_{c,z} R_{500}^3$, where $\rho_{c,z} = 3 H_z^2 / (8 \pi G)$ is the critical density of the universe at the cluster's redshift, $G$ is the gravitational constant, and $R_{500}$ is the radius of the sphere within which the average total mass density is 500 $\rho_{c,z}$.}
$> 2 \times 10^{14}~M_{\odot}$), together with a sample of the highest-mass objects in the Universe (Tier-2: $z<0.6$; ${M_{500}}>7.25 \times 10^{14}~M_{\odot}$). 
The project acquired X-ray exposures of uniform depth that ensure a detailed mapping of the thermodynamic properties in the cluster volume where the non-thermal plasma is present. Therefore, the CHEX-MATE cluster sample is the best choice for a systematic and statistical analysis of cluster thermal components.\par
In the radio band, the LoTSS \citep{LoTSS2017} is a deep, $120 - 168$ MHz radio survey, that produces high resolution ($\sim 6^{\prime\prime}$) and high sensitivity ($\sim 100~\mu\rm{Jy~ beam^{-1}}$) images of the northern sky. It had its first data release (LoTSS DR1) in 2019 \citep[][released area  $\sim 400$ $\rm{deg^2}$ ]{LoTSS2019} and the second release in 2022 \citep{Shimwell2022} providing images and radio catalogues for $\sim 5,700~ \rm{deg^2}$ of the Northern sky. 
One of the main goals of this survey is to find new diffuse radio sources inside galaxy clusters, such as giant radio halos, to determine their origin and to test theoretical and numerical models. 
Thanks to its high sensitivity to diffuse sources at low frequencies, the LoTSS allows us to perform detailed studies of radio halos in the frequency domain where these objects appear brighter because of their steep synchrotron spectra.\par
\begin{figure}[ht]
    \centering
    \includegraphics[width=\linewidth]{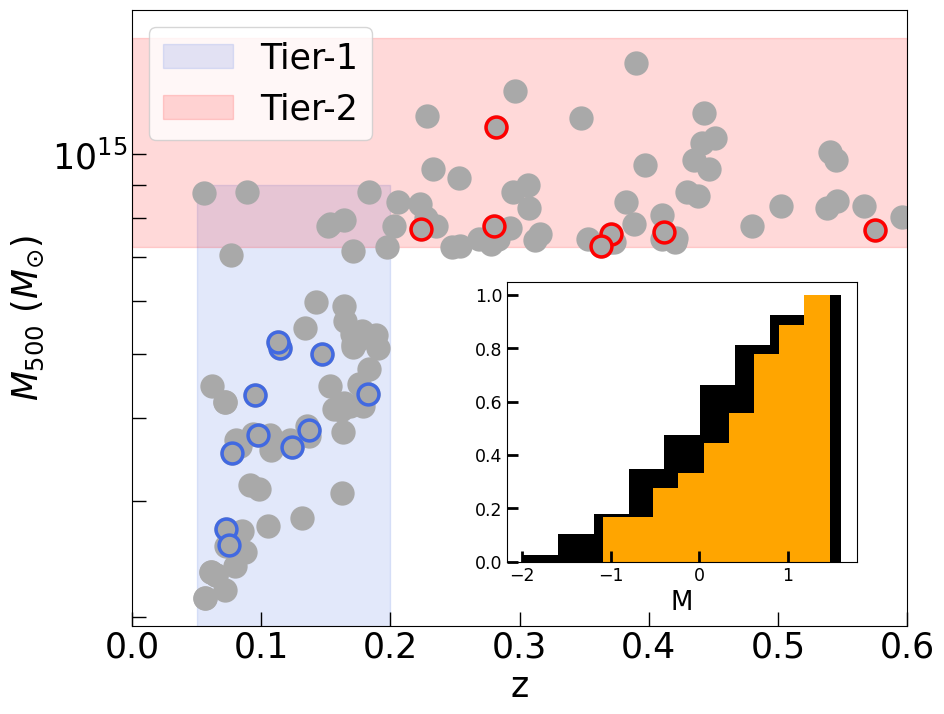}
    \includegraphics[width=\linewidth]{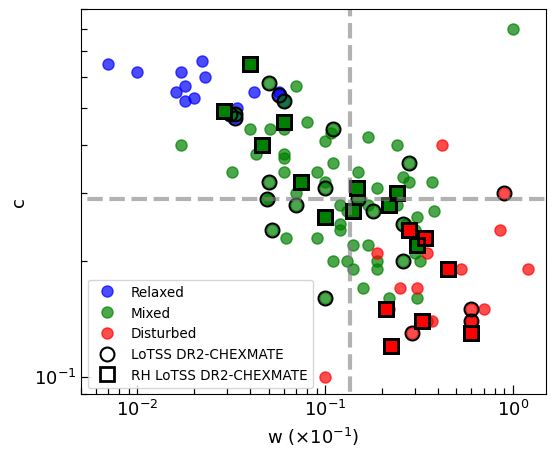}    
    \caption{Comparison between the subsample of 18 clusters considered here and the full CHEX-MATE sample.
    Top: $M_{500} - z$ distribution of the CHEX-MATE sample, with highlighted (circles) the LoTSS DR2 clusters with diffuse emission and the Tier-1 and Tier-2 subsamples. In the inset are reported the CDFs of the M parameter of the whole CHEX-MATE sample (black) and the one to the targets considered here (orange).
    Bottom: $c-w$ plot of the whole CHEX-MATE sample, with highlighted objects with (empty black squares) and without (empty black circles) diffuse halo emission from the LoTSS DR2. The targets are coloured according to the classification made by \cite{Campitiello2022}. Dashed grey lines indicate the median $c$ and $w$ values of the sample.}
    \label{fig:18-CHXM}
\end{figure}

Of the 82 Northern CHEX-MATE targets, 40 are in the LoTSS DR2 sky area, and 18 display diffuse halo emission and are the ones we consider here (\citealt{Botteon2022}).
Since we selected our targets from the CHEX-MATE sample, we can exploit, on one side, Planck measurements of $M_{500}$ (derived from the $Y_{SZ}$ signal calibrated on the $M_{hydrostatic}-Y_{X}$ relation\footnote{$Y_X$ is the product of the ICM gas mass within $R_{500}$, and $T_X$, the spectroscopic temperature measured in the [0.15–0.75] $R_{500}$ aperture; while the SZ signal, $Y_{SZ}$, is proportional to the thermal energy content of the ICM along the line of sight.}; see \citealt{Arnaud21} for more details) and, on the other, the information on the dynamical activity of the targets. In fact, \cite{Campitiello2022} analysed all the X-ray data of the CHEX-MATE sample, providing a systematic and statistical evaluation of the clusters' dynamical state.
{They estimated the light concentration ($c$),
\begin{equation}
    c = \frac{NC(r<0.15R_{500})}{NC(r<R_{500})},
\end{equation}
the centroid shift ($w$),
\begin{equation}
    w = \frac{1}{R_{500}} \left [ \frac{1}{N-1}  \sum_i (\Delta_i - \Bar{\Delta})^2 \right]^{1/2},
\end{equation}
%
and M, which combines the morphological indicators presented in \cite{Campitiello2022} by summing the deviations of each parameter from the mean of its distribution in units of standard deviation (\citealt{Rasia2013}).\\
The selected CHEX-MATE-LoTSS DR2 objects (see Tab.~\ref{tab:sample_18}) span a broad range of redshift ($0.072 \leq {\rm z} \leq 0.575 $) and mass ($ 2.57 \times 10^{14}~M_{\odot} < M_{500} <  11.00 \times 10^{14} ~ M_{\odot}$)\footnote{\citep[all the derived masses are taken from the MMF3 Planck catalogue ][]{Planck2016}} as shown in the top panel of Fig.~\ref{fig:18-CHXM}.
They belong to both CHEX-MATE Tier-1 and Tier-2 subsamples, uniformly sampling the whole mass and redshift range up to $z\sim0.4$. We also note that two targets are located beyond this redshift value.
Furthermore, these clusters span a rather wide range in terms of cluster dynamical status. As shown in Fig.~\ref{fig:18-CHXM}, they cover the majority of the dynamical range of CHEX-MATE, ranging from very disturbed objects to mildly-relaxed ones. 
This is evident from both the comparison between the cumulative distribution functions (CDFs) of the M parameter within CHEX-MATE and the 18 selected objects (Fig.~\ref{fig:18-CHXM} top), and from the location of the latter in the $c-w$ plane (Fig.~\ref{fig:18-CHXM} bottom), usually used in the past to identify and separate clusters displaying radio halo emission (e.g. \citealt{Cassano2010}).
General information about the
analysed targets is listed in Table~\ref{tab:sample_18}.
In the bottom pane of Fig.~\ref{fig:18-CHXM}, we highlight the position in the $c-w$ plot of the considered 18 targets within the CHEX-MATE sample.}
\begin{table*}[ht]
    \centering
    \begin{tabular}{lcccccc}
    \toprule
    Name & z & $M_{500} ~ (10^{14} M_{\odot})$ & $R_{500} ~ {\rm (arcmin)}$ & $c$ & $w ~(10^{-1})$ & M \\
    \midrule
    \midrule
    PSZ2G031.93+78.71 & 0.072 & $2.72 \pm 0.24$ & 11.64 & $0.31^{+0.07}_{-0.07}$ & $0.148^{+0.001}_{-0.010}$ & -0.47 \\
    PSZ2G040.58+77.12 & 0.075 & $2.63 \pm 0.22$ & 11.08 & $0.30^{+0.01}_{-0.01}$ & $0.240^{+0.11}_{-0.13}$ & 0.56 \\
    PSZ2G046.88+56.48 & 0.115 & $5.31 \pm 0.23$ & 9.40 & $0.14^{+0.04}_{-0.04}$ & $0.330^{+0.17}_{-0.18}$ & 1.41 \\
    PSZ2G048.10+57.16 & 0.078 & $3.59 \pm 0.21$ & 11.90 & $0.12^{+0.02}_{-0.02}$ & $0.225^{+0.004}_{-0.010}$ & 1.03 \\
    PSZ2G049.32+44.37 & 0.097 & $3.67 \pm 0.26$ & 9.87 & $0.28^{+0.06}_{-0.07}$ & $0.220^{+0.05}_{-0.07}$ & 0.08 \\
    PSZ2G053.53+59.52 & 0.113 & $5.85 \pm 0.23$ & 9.58 & $0.22^{+0.03}_{-0.04}$ & $0.310^{+0.04}_{-0.06}$ & 0.98 \\
    PSZ2G055.59+31.85 & 0.224 & $7.78 \pm 0.31$ & 5.99 & $0.46^{+0.05}_{-0.05}$ & $0.060^{+0.02}_{-0.01}$ & -0.46 \\
    PSZ2G056.77+36.32 & 0.095 & $4.38 \pm 0.20$ & 10.53 & $0.49^{+0.06}_{-0.06}$ & $0.029^{+0.004}_{-0.001}$ & -0.81 \\
    PSZ2G066.41+27.03 & 0.576 & $7.70 \pm 0.53$ & 2.88 & $0.15^{+0.01}_{-0.02}$ & $0.210^{+0.04}_{-0.06}$ & 0.74 \\
    PSZ2G077.90-26.63 & 0.15 & $5.06 \pm 0.26$ & 7.46 & $0.40^{+0.06}_{-0.07}$ & $0.046^{+0.002}_{-0.010}$ & -1.03 \\
    PSZ2G083.29-31.03 & 0.412 & $8.27 \pm 0.44$ & 3.66 & $0.27^{+0.03}_{-0.03}$ & $0.140^{+0.03}_{-0.05}$ & 0.36 \\
    PSZ2G107.10+65.32 & 0.280 & $8.22 \pm 0.28$ & 5.00 & $0.19^{+0.03}_{-0.04}$ & $0.450^{+0.08}_{-0.1}$ & 0.66 \\
    PSZ2G111.75+70.37 & 0.183 & $4.34 \pm 0.33$ & 5.88 & $0.13^{+0.03}_{-0.03}$ & $0.600^{+0.2}_{-0.2}$ & 1.47 \\
    PSZ2G113.91-37.01 & 0.371 & $7.58 \pm 0.55$ & 3.96 & $0.23^{+0.03}_{-0.04}$ & $0.340^{+0.08}_{-0.09}$ & 0.74 \\
    PSZ2G143.26+65.24 & 0.363 & $7.65 \pm 0.43$ & 3.97 & $0.24^{+0.03}_{-0.03}$ & $0.280^{+0.05}_{-0.06}$ & 0.87 \\
    PSZ2G179.09+60.12 & 0.137 & $3.84 \pm 0.33$ & 7.26 & $0.65^{+0.06}_{-0.07}$ & $0.040^{+0.02}_{-0.02}$ & -1.09 \\
    PSZ2G186.37+37.26 & 0.282 & $11.00 \pm 0.37$ & 5.57 & $0.32^{+0.03}_{-0.04}$ & $0.074^{+0.001}_{-0.013}$ & -0.20 \\
    PSZ2G192.18+56.12 & 0.124 & $3.62 \pm 0.30$ & 7.80 & $0.26^{+0.07}_{-0.08}$ & $0.100^{+0.03}_{-0.05}$ & 0.19 \\
    \bottomrule
    \end{tabular}
    \caption{Main properties of the 18 clusters of the crossmatched CHEX-MATE--LoTSS DR2 sample.}
    \label{tab:sample_18}
\end{table*}
\par
%
%
We imposed a cut in redshift at $z < 0.4$, leaving us with 16 CHEX-MATE radio halos observed by the LoTSS DR2. Contrary to the order of magnitude range in mass, the redshift range is homogeneously sampled only up to $z=0.4$.
This cut also mitigates the effect of the k-correction, $(1+z)^{\alpha-1}$, due to the assumed spectral index that is poorly known in that redshift range. We also note that, for the same reasons, the same cut in redshift has been adopted also by \cite{Cuciti2023} for a wider sample of targets.\par
In order to gain higher statistics when constraining the $P_{\nu} - M$ relation of our sample (Sect.~\ref{sec:model}), we will also exploit the public LoTSS DR2 observations of radio halos presented by \cite{Botteon2022} and discussed in \cite{Cuciti2023}\footnote{\url{https://lofar-surveys.org/radiohalo.html}. }.
From the whole sample of PSZ2 clusters in the LOFAR DR2 area, \cite{Cuciti2023} performed the analysis on a sample of 61 objects lying above the Planck 50\% completeness line\footnote{They considered the Planck mass estimates (\citealt{Planck2016}), i.e. the same adopted for the CHEX-MATE targets.}. 
With our further cut on the redshift of $z<0.4$, we will consider a sample of 43 objects, which includes all the CHEX-MATE targets considered in this work apart from A2409. 
The authors did not consider this target in their work as it was not possible to obtain a robust radio power estimation through the fitting process. However,  we are neither considering any of their fit results nor fitting the profile of single targets but we are just extracting the observed radial profile. Hence, we will include this object in our analysis.\\
A summary of the samples used in this work is presented in Table~\ref{tab:samples}.\par
\begin{table*}[ht!]
     
    \centering
    \begin{tabular}{lccl}
         \toprule
         Sample & Total objects & Clusters at $z<0.4$ & Purpose \\
        \midrule
        \midrule
         LoTSS DR2 RH & 61 & 43 & Constrain the $P_{\nu}-M$ relation\\
         LoTSS~DR2 $\cap$ CHEX-MATE & 40 & 38 & / \\
         LoTSS~DR2 $\cap$ CHEX-MATE RH & 18 & 16 & Derive scaling relations for halo profiles \\
         \bottomrule
    \end{tabular}
    \caption{Summary of the samples considered in this work.}
    \label{tab:samples}
\end{table*}
%
%
The non-thermal analysis of CHEX-MATE targets we propose here can be seen in analogy with the more common spatially resolved studies made for the thermal component (see Sect.~\ref{sec:intro}).
Therefore, this study fits well in the wider project of a detailed characterisation of the thermal--non-thermal connection in clusters that we have started in \cite{Balboni2024} and will extend in forthcoming works.
By combining the sample properties of CHEX-MATE and the deep, homogeneous observations of the LoTSS, here we aim to perform a spatially resolved, radio analysis on a representative sample of clusters. 
\subsection{Radio data reduction}\label{sec:datared}

In the following, we report an overview of the calibration and imaging procedures applied to the radio data. The complete description of the reduction process is presented in \cite{Botteon2022}.\par
LoTSS pointings are typically obtained with an integration time of 8 hr and in the 120-168 MHz frequency range. The collected data were processed with fully automated pipelines developed by the LOFAR Surveys Key Science Project team in order to correct for direction-independent and direction-dependent effects \citep{vanWeeren2016,Williams2016,deGasperin2019,Tasse2021,vanWeeren2021}. 
After scaling the flux density values to the \cite{Roger1973} scale, the imaging was done with WSClean v2.8 \citep{Offringa2014}. 
{  Since we are particularly interested in the diffuse halo emission, we retrieved the images from the LoTSS DR2/PSZ2 website\footnote{\url{https://lofar-surveys.org/planck\_dr2.html}} obtained with a taper of 100 kpc at the cluster redshift and with the further subtraction of emission from sources with a physical size smaller than 250 kpc from the visibilities.
We additionally masked out from the resulting images those regions with residual contaminating sources (both compact and extended) or which showed signs of poor subtraction, as also made in \cite{Botteon2022}}.

\section{Data analysis}\label{sec:analysis}

\begin{figure}
    \centering
    \includegraphics[scale=0.26]{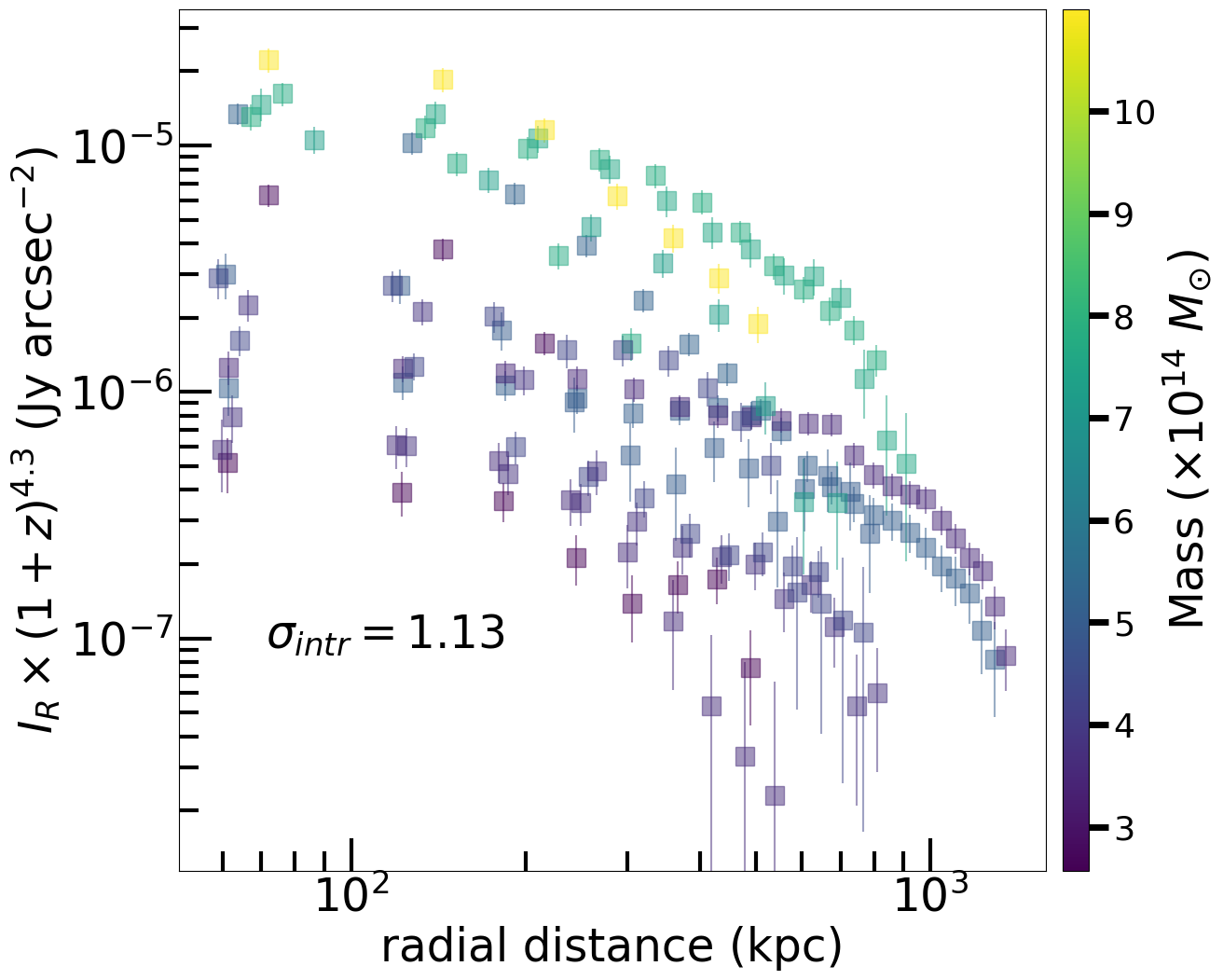}    \caption{Rest-frame radio surface brightness profiles of the 16 targets at $z<0.4$ considered in this work. In the legend, it is reported the global scatter of the profiles when fitted with an exponential model (see Sect.\ref{sec:model} and \ref{sec:prof_scaling}).}
    \label{fig:radio_prof}
\end{figure}

Starting from the radio images described in Sect.~\ref{sec:datared}, we derive the radial profile for each radio halo as described in \cite{Balboni2024}. We firstly fit the radio halo emission with an exponential function by means of Halo-Flux Density CAlculator \citep[Halo-FDCA,][]{HFDCA-2021} to obtain a good estimate of the halo shape (circular or elliptical based on what was reported in \citealt{Botteon2022}) and centre.
Since Halo-FDCA fits the radio halo profile with an exponential model, $I \propto \exp\left(-\frac{r}{R_e}\right)$ with $R_e$ the effective radius, it allows us to determine also the expected extension of the radio halo as usually considered in literature $\sim 2.5 - 3 R_e$ (e.g. \citealt{Bonafede2017}).
{The exponential model has been proved to provide a good description of radio halo profiles with few free parameters, making its use the common practice when characterising the radial extension of halos (e.g. \citealt{Murgia2009,Botteon2022}). In addition, Halo FDCA provides a high flexibility of the exponential function, improving the model description of the various shapes of radio halos.} 
We found that, in most of the cases, the halo signal is above the $1\sigma$ noise level for $ r < 2.5 R_e$. So we extract the radio profile up to $2.5 R_e$ using the halo centre previously defined, adopting circular or elliptical annuli, depending on what has been adopted during the halo fitting, and choosing the width of the annuli to be half of the Full-Width at Half Maximum of the image beam \citep{Cuciti21-1}.
In Fig.\ref{fig:radio_prof}, we report the rest-frame, 1-D radial profiles of the studied halos and the value of their intrinsic scatter with respect to the best-fit exponential model (see Sect.\ref{sec:model} in the legend. 
For each annulus, we compute the mean of the radio brightness and associated an error of $\delta I = \sqrt{ (0.1 \times I )^2 + ({ {\rm \sigma_{RMS}} / \sqrt{N_{beam}}})^2 }$, where $\rm N_{beam}$ is the number of beams in the annulus and $I$ the brightness value and the factor 0.1 takes into account the uncertainty of the flux density scale (\citealt{Shimwell2022}).\par
%
Additionally, Halo-FDCA provides the values for the radio power computed as
\begin{equation}\label{eq:power}
    P_{\nu} = 4 \pi S_{\nu} D_L^2 (1+z)^{\alpha - 1},
\end{equation}
where $z$ is the redshift, $D_L$ is the luminosity distance and the spectral index $\alpha$ is assumed to be 1.3 for all the radio halos \citep[we refer the reader to Appendix~\ref{appendix:table_RH} or][for a full list of all the radio power of the LoTSS DR2 objects]{Botteon2022}.
%
%
%
\subsection{The radio scaling-law model}\label{sec:model}

In this Section, we derive a model that links the observed $P_{\nu}-M$ relation (as presented and discussed in recent literature, e.g.\citealt{Cassano2007,vanWeeren2021,Cuciti21-2,Duchesne2021-EoR,George2021,Cuciti2023}), accounting also for the redshift dependence, to the spatially resolved radio halo properties. In this way, we will be able to obtain predictions on the expected halo profile re-scaling, in both mass and redshift, given the relation between $P_{\nu}$ and $M$.
%
The flux density $S_{\nu}$ can be computed as the integral over the whole radio halo solid angle of the surface brightness $I_{\nu}$, resulting in
\begin{equation}    
    S_{\nu} \propto I_{\nu} \theta^2,
\end{equation}
with $\theta = R_H/D_{\theta}$ being the ratio between the halo radius $R_H$ (i.e. its linear size) and the angular diameter distance $D_{\theta}$. Then, including this in Eq.~\ref{eq:power}, we get,
\begin{equation}
    P_{\nu} \propto I_{\nu} R_H^2 \left(\frac{D_L}{D_{\theta}} \right)^2 (1+z)^{\alpha - 1}.
\end{equation}
%
Since the non-thermal ICM component is tightly related to the cluster dynamical history, the latter, in simple terms, can be considered the main driver for radio halo emission. Consequently, one might reasonably expect that, as in the case of thermal ICM, a relation between the halo size and cluster properties is present.
Thus, to quantify the radio halo size scaling, we will consider the general expression $R_H \propto M^{\beta_M} E_z^{\beta_z}$ (see more on this in Sec.~\ref{sec:no_RH-M}). 
Using the cosmological relation $D_L/D_{\theta} = (1+z)^2$, we can write
\begin{equation}\label{eq:P-I_prop}
P_{\nu} \propto I_{\nu} \, M^{2 \beta_M} \, E_z^{2 \beta_z} \,(1+z)^{\alpha +3},
\end{equation}
which reads as $P_{\nu} \propto I_{\nu} \, M^{2/3} \, E_z^{-4/3}$ for the case of $R_H \propto R_{\Delta}$ (where $\Delta$ is the overdensity with respect to the critical density $\rho_c$), with $\beta_M=1/3$ and $\beta_z=-2/3$.\\
%
{One can then consider a power-law relation between the total radio power and the mass 
\begin{equation}\label{eq:pm}
P_{\nu} \propto (M)^{\alpha_{M}},
\end{equation}
or, more generally, also account for an evolution of $M$ with the redshift as follows:
\begin{equation}\label{eq:pmz}
P_{\nu} \propto (M \, E_z)^{\alpha_{M,z}}.
\end{equation}}
Combining Eq.~\ref{eq:pmz} with Eq.~\ref{eq:P-I_prop}, we can derive the expected dependence of the surface brightness upon the halo's mass and redshift:
\begin{equation}\label{eq:prop_relation}
    I_{\nu} \propto M^{\alpha_{M,z} - 2 \beta_M} \, E_z^{\alpha_{M,z} -2\beta_z} \, (1+z)^{-(3+\alpha)}.
\end{equation}
If we insert in our equations an exponential model to reproduce the radio halo profile, we get
\begin{align}\label{eq:rescaling_IR}
\log \left( \frac{I_{\nu}(r)}{(1+z)^{-(3+\alpha)}} \right ) =& ~  {\rm log}(I_0) + A \left( \frac{r}{R_{500}} \right) \\
&+ \gamma_M \log \left( \frac{M_{500} } {10^{14} M_{\odot} }\right) \nonumber + {\gamma_{z}} \log E_z ,
\end{align}
where $r$ is the radial distance, while $\gamma_M$ and $\gamma_z$ are the parameters used to describe the dependencies of the radial profiles on mass and redshift, respectively.
%
Given these models, we expect the following relations to be satisfied:
\begin{align}\label{eq:rescaling_eqs}
\gamma_M = & \, \alpha_{M,z} -2 \beta_M \; (=\alpha_{M,z} -2/3, \text{for the case} \, R_H \propto R_{\Delta}) \nonumber \\
\gamma_{z} =& \, \alpha_{M,z} - 2\beta_z \; (=\alpha_{M,z} +4/3). 
\end{align}
{ Therefore, for each quantity $Q$ (in this case $M, ~z$) we end up with a set of best-fit variables, {$\gamma_Q$}, obtained by fitting Eq.~\ref{eq:rescaling_IR} to the observed profiles and one set of predicted ones, {$\gamma_{Q,p}$}, from Eq.~\ref{eq:rescaling_eqs}.}
In this way, we can test whether the scaling presented in this Section holds. In particular, we can test if and how the observed $P_{\nu} -M$ relation can be translated into scaling of the radio halo profiles and what are the main drivers of such scaling.\\
We note that, by comparing the predicted and measured $\gamma$, and the mass and redshift dependence of the $P_{\nu}-M$ relation, we can also infer which values of $\beta_Q$ best accommodate the profiles best-fit results (i.e. from Eq.~\ref{eq:rescaling_eqs}, $\beta_M = (\alpha_{M,z}-\gamma_M)/2$ and $\beta_z = (\alpha_{M,z}-\gamma_z)/2$).\par
We report a summary of the defined variables used to indicate the mass and redshift dependencies in Table~\ref{tab:vars}.
\begin{table*}[ht]
    
    \centering
    \begin{tabular}{lll}
        \toprule
        Variable & Definition & Equation \\
        \midrule
        \midrule
        $\alpha_M$ & global mass scaling &$P_{\nu} \propto (M)^{\alpha_M}$  \\        
        $\alpha_{M,z}$ & global mass and redshift scaling &$P_{\nu} \propto (M ~ E_z)^{\alpha_{M,z}}$  \\
        $\gamma_{M}$ ($\gamma_z$) & profiles, best-fit mass (redshift) scaling &$I_{\nu} \propto M^{\gamma_M} E_z^{\gamma_z}$ \\
        $\gamma_{M,p}$ ($\gamma_{z,p}$) & profiles, predicted mass (redshift) scaling &$I_{\nu} \propto M^{\gamma_{M,p}} E_z^{\gamma_{z,p}}$ \\
        $\beta_M$ ($\beta_z$) & mass (redshift) scaling of $R_H$  &$R_H \propto M^{\beta_M} E_z^{\beta_z}$ \\
        \bottomrule
    \end{tabular}
    \caption{Summary of the variables defined to quantify the mass and redshift scalings throughout this work.}
    \label{tab:vars}
\end{table*}
{
\subsection{The $P_\nu -  M$ relation and model predictions}\label{sec:m_only}

We start by considering only the mass scaling of $P_{\nu}$ as in Eq.~\ref{eq:pm}, that is without considering any mass evolution with the redshift ($\alpha_{M,z} \equiv \alpha_M$):
\begin{equation}\label{eq:P-M}
    \log P_{150MHz} =  {\rm log}(P_0) + \alpha_M \, \log \left( \frac{M_{500} } {10^{14} M_{\odot} }\right).
\end{equation}
%
As explained in Sect.~\ref{sec:sample}, to determine $\alpha_M$, we make use of the public LoTSS DR2 cluster sample of radio halos (\citealt{Botteon2022,Cuciti2023}).
Both the values of $M_{500}$ and the radio power at 150 MHz ($P_{150{\rm MHz}}$, computed as in Eq.~\ref{eq:power}), are provided by \cite{Botteon2022}.
We fit a power-law relation between the total radio power and the mass ($M_{500}$) using the \texttt{linmix} package \citep{Kelly2007}. The Bayesian approach of \texttt{linmix} approximates the independent variable distribution with a Gaussian function, mitigating the possible selection effects in our cluster sample. This approach is similar to the one of \cite{Cuciti2023} where the authors used a log-normal function to describe the cluster mass distribution.
The best-fit values are reported in Table~\ref{tab:P-M} and in the left panel of Fig.~\ref{fig:P-M}. 
For comparison, we made the fit also on the full sample (all 61 targets with no redshift cut) as made in \cite{Cuciti2023}. We find that the recovered mass slope is $\sim 3.44$ which is in good agreement with the BCES(Y|X) method ($3.45\pm0.44$) presented in \cite{Cuciti2023}.
\begin{table*} 
    \centering
    \caption{Radio power best-fit rescaling parameters, using the objects classified, by \cite{Botteon2022} and \cite{Cuciti2023}, as a radio halo and candidate radio halo and located at $z<0.4$. }
    \begin{tabular}{llllllllll}
        \toprule
            & ${\rm log}(P_0)$ & $\sigma_{int}$ & $\alpha_{M,z} ~(\alpha_M)$\\
        \midrule
        \midrule
            $P_{150MHz} \propto M^{\alpha_M}$ & $22.76 \pm 0.28$ & $0.35 \pm 0.05$ & $3.02 \pm 0.39$ \\
            $P_{150MHz} \propto \left (M ~ E_z \right)^{\alpha_{M,z}}$ & $22.81 \pm 0.26$ & $0.34 \pm 0.04$ & $2.76 \pm 0.35$ \\
            %
        \bottomrule
    \end{tabular}
    \label{tab:P-M}
\end{table*}
\begin{figure*}[ht!]
    \centering
    \includegraphics[scale=0.35]{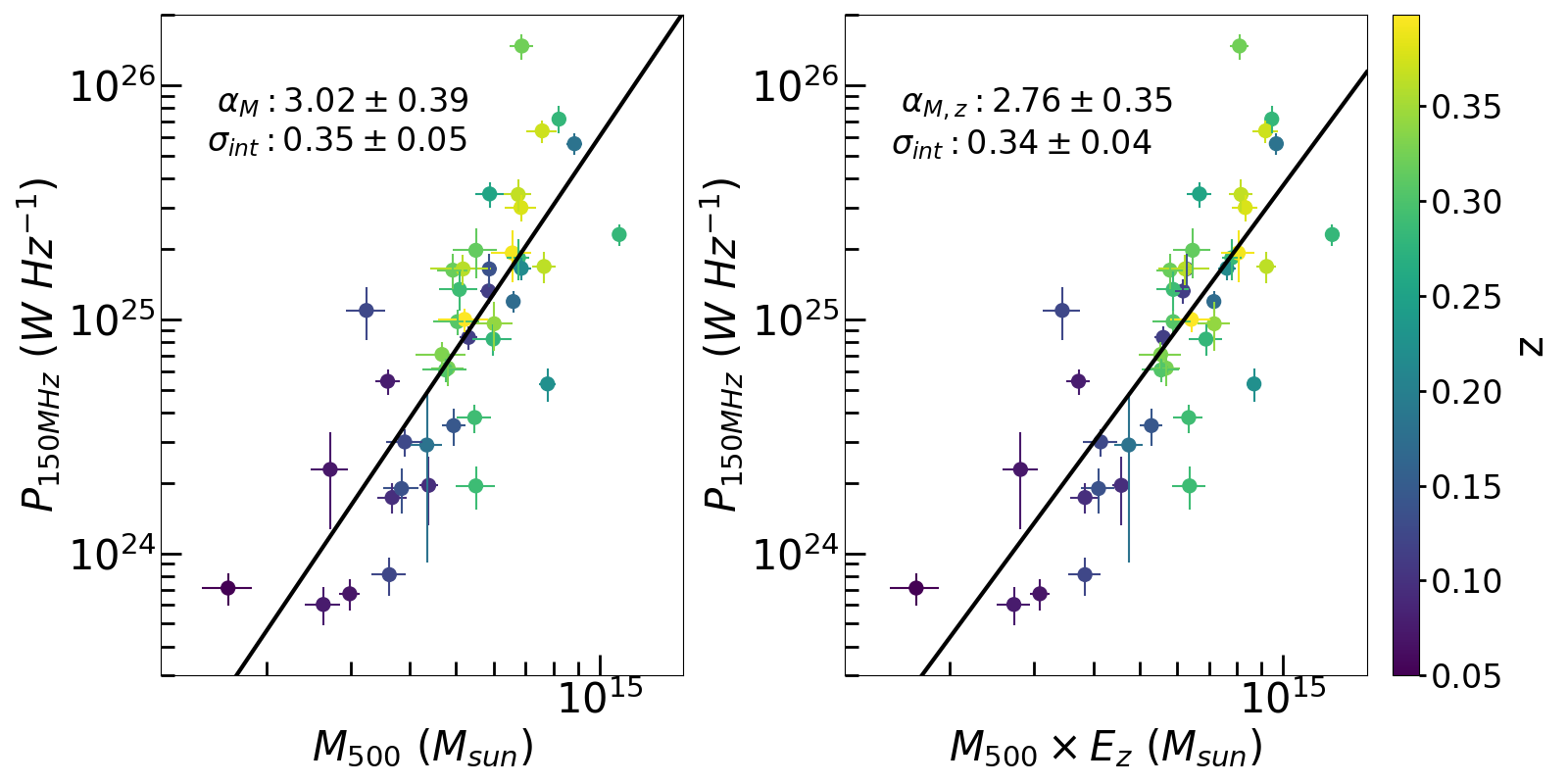}
    \caption{Power mass relation for the radio halos, including candidates, of the LoTSS DR2 at $z<0.4$. In the two panels, we present the different scaling proposed. Left: No redshift dependence. Right: Equal mass and redshift dependence.}
    \label{fig:P-M}
\end{figure*}\par
According to Eq.~\ref{eq:rescaling_eqs}, a best-fit $\alpha_M=3.02 \pm 0.39$ would imply a mass rescaling for the radial profiles as $\gamma_{M, p} = 2.35 \pm 0.39$, when $R_H \propto R_{\Delta}$.\par
As a second step, we include a proper redshift evolution through the term $E_z$ (see Eq.~\ref{eq:pmz}) in the form
\begin{equation}\label{eq:P-Mz}
    \log P_{150MHz} = {\rm log}(P_0) +\alpha_{M,z} \, \log \left( \frac{M_{500} } {10^{14} M_{\odot}} ~ E_z \right) ,
\end{equation}
and evaluate its impact on the scaling of the radio surface brightness profiles as expected from Eq.~\ref{eq:rescaling_eqs}.
%
%
%
%
We fit the radio power as a function of mass and redshift keeping the same re-scaling for $M_{500}$ and $E_z$ .
We report the best-fit values in Table~\ref{tab:P-M} and in the legend panel of Fig.~\ref{fig:P-M} (right).\\
In the case of $R_H \propto R_{\Delta}$, from Eq.~\ref{eq:rescaling_eqs}, we derive the expected dependencies of the resolved quantities as: $\gamma_{M, p} =
2.10 \pm 0.35$ and $\gamma_{z, p} =
4.10 \pm 0.35$.\par
%

\subsection{Profiles scaling and model comparison}\label{sec:prof_scaling}
We can now test the predictions made in the previous Section by searching for the best mass rescaling of the observed radial profiles using  Eq.~\ref{eq:rescaling_IR} and comparing the results with the expected values.\par
For the case with no redshift evolution term in the $P_{\nu} - M$ relation, we fix $\gamma_z$ to 0, in the case of no dependence with redshift, and to 4/3, in the case of the self-similar evolution. We then fit the radial profiles using Eq.~\ref{eq:rescaling_IR}, assuming a spectral index $\alpha = 1.3$ \citep{Botteon2022} and leaving as free parameters the normalisation and $\gamma_M$. The results of the fits are reported in Table~\ref{tab:Ir_fit}.
\begin{table*}
    \centering
    \caption{Radial profiles' best-fit rescaling parameters.}
    \label{tab:Ir_fit}
    \begin{tabular}{llllllllllllll}
        \toprule
            Quantity & ${\rm log}(I_0)$ & $A$ & $\sigma_{int}$ &  $\gamma_M$ &  $\gamma_z$ \\
            \midrule
            \midrule
            $I_{150MHz} \propto M^{\gamma_M}$ & $-8.43 \pm 0.33$ & $-0.70 \pm 0.16$ & $0.22 \pm 0.05$ & $3.68 \pm 0.43$ & 0 (fixed)  \\
            $I_{150MHz} \propto M^{\gamma_M} ~ E_{z,{\rm fixed}}^{4/3}$ & $-8.30 \pm 0.30$ & $-0.71 \pm 0.15$ & $0.24 \pm 0.04$ & $3.41 \pm 0.38 $& 4/3 (fixed) \\
            $I_{150MHz} \propto M^{\gamma_M} ~ E_{z,{\rm fixed}}^{\alpha_{M} + 4/3}$ & $-8.00 \pm 0.33$ & $-0.72 \pm 0.15$ & $0.31 \pm 0.04 $&$ 2.85 \pm 0.4$3 & 4.10 (fixed) \\
            $I_{150MHz} \propto (M E_z)^{\gamma_{M}}$ & $-7.98 \pm 0.31$ & $-0.78 \pm 0.15$ & $0.26 \pm 0.04$ & $2.88 \pm 0.33$ & $\gamma_M$ \\
        \bottomrule
    \end{tabular}
\end{table*}
For the case of no redshift evolution, $\gamma_z =0$, we find $\gamma_{M}=3.68 \pm 0.43$, while for the $R_H \sim R_{\Delta}$ case $\gamma_{M} = 3.41 \pm 0.38$. Both values exceed the predicted $\gamma_{M, p} = 2.35 \pm 0.39$, with  consistency only at $2.3 \sigma$ and $1.9 \sigma$, respectively.\par
When we consider also the $E_z$ term in the $P_{\nu}-M$ relation, we have an additional contribution to $\gamma_z$ (=$\alpha_{M,z} - 2 \beta_z$) in Eq.~\ref{eq:rescaling_IR}. 
However, due to the degeneracy between the cluster mass and redshift (although weakened by the imposed redshift cut), and the small number of CHEX-MATE-LoTSS DR2 clusters available, we are not able to provide constraints for both mass and redshift separately. Therefore, we will focus only on the mass scaling, $\gamma_M$, of the profiles. 
We fix the redshift dependence to the expected one from the integrated quantity analysis, i.e. $\gamma_z = 4.10$, and fit the radial profiles using Eq.~\ref{eq:rescaling_IR}, leaving the normalisation and $\gamma_M$ as free parameters. The results of the fit are reported in Table~\ref{tab:Ir_fit}.
A value of $\gamma_M = 2.85 \pm 0.43$ is higher than the expected $2.10 \pm 0.35$ with a consistency within $1.4 \sigma$.\par
A different approach we take is to rescale the radial profiles using the expected values derived from the $P_{150MHz}-M_{500}$ relation.
\begin{figure}[ht!]
    \centering
    \includegraphics[scale=0.265]{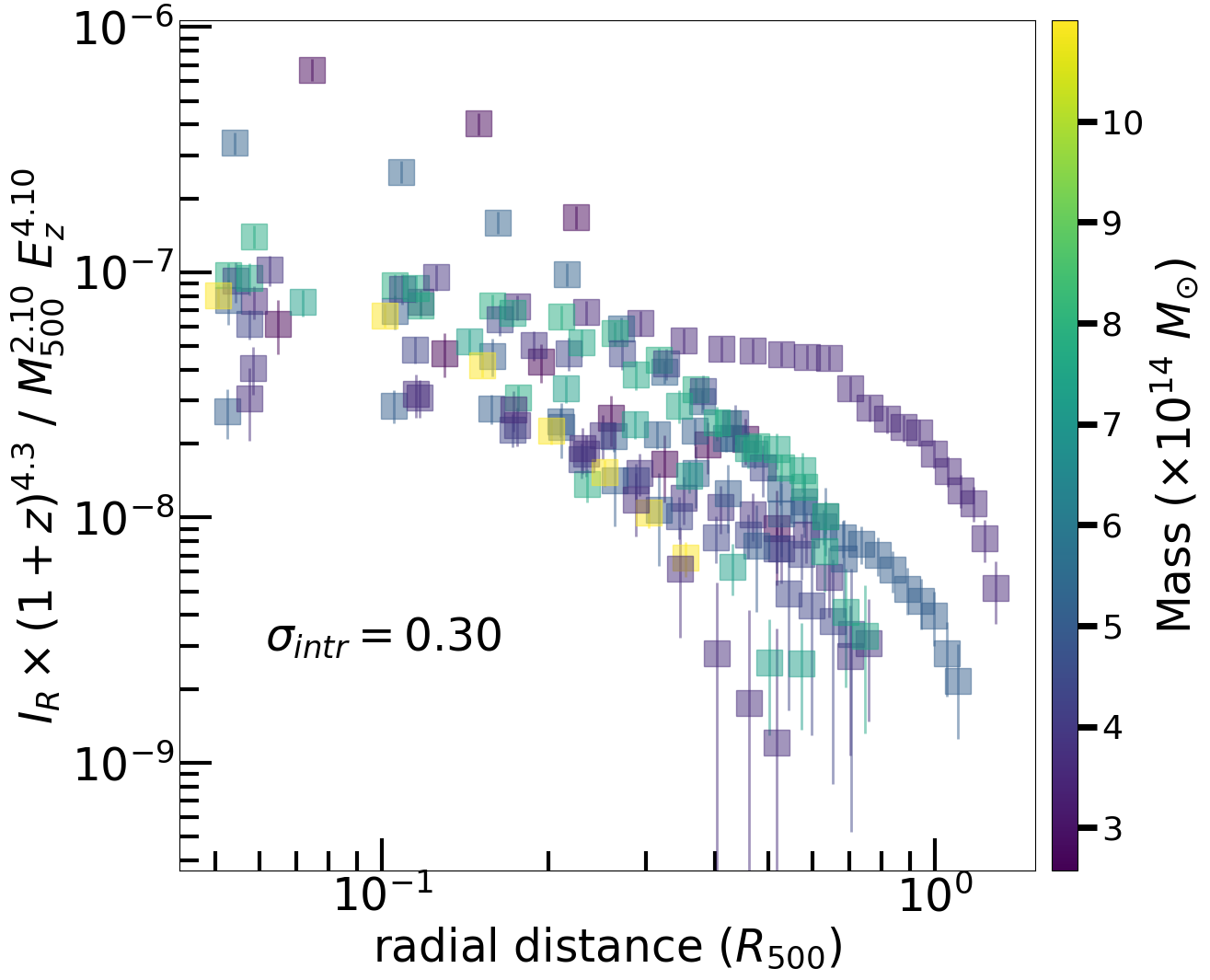}
    \includegraphics[scale=0.265]{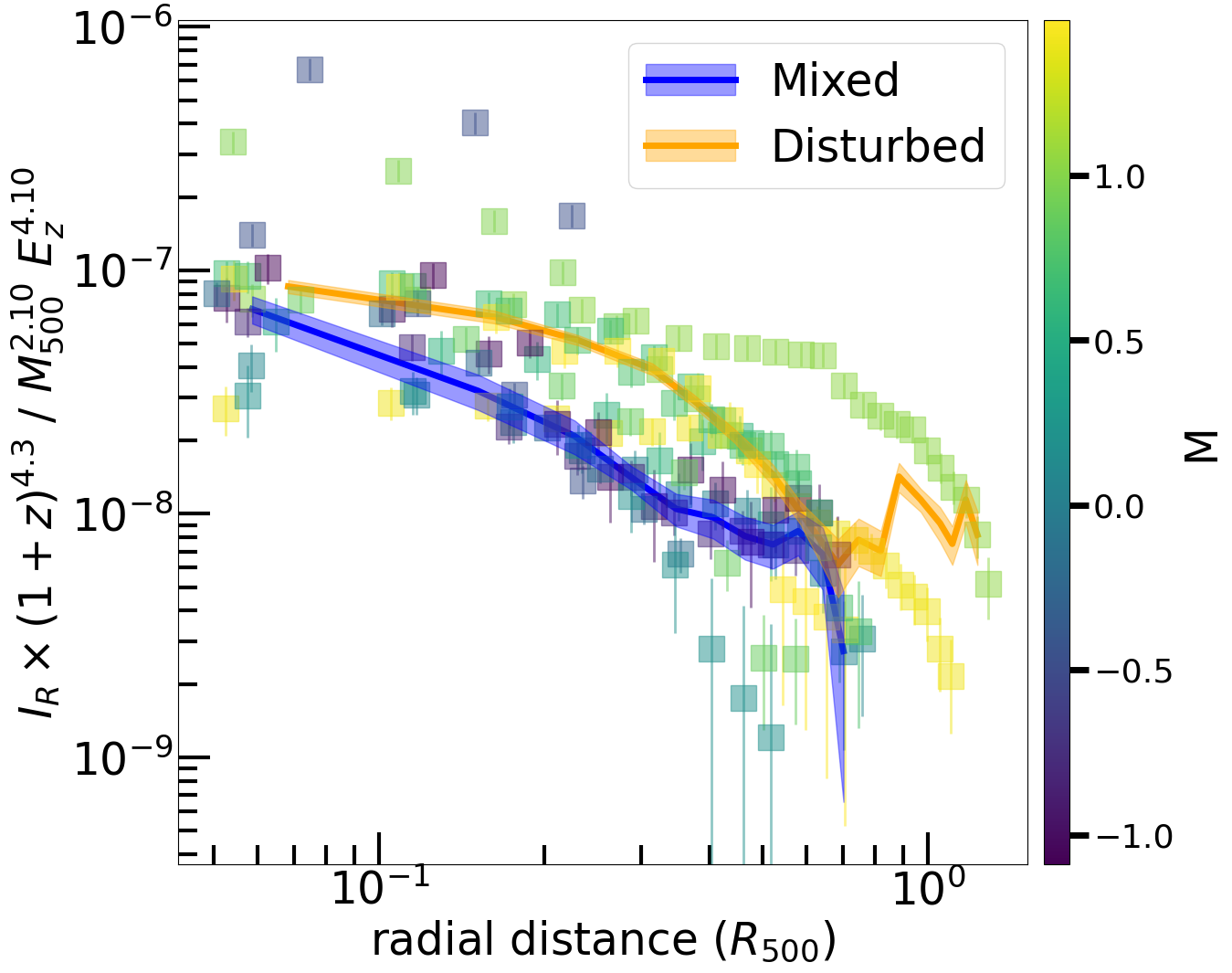}
    \caption{Scaled, rest-frame, radial profiles of the considered halos and their scatter.
    Top: Scaled profiles assuming the expected scaling as described in Eq.~\ref{eq:rescaling_IR-2}. Bottom: Same plot, but with objects colour-coded according to their morphological M parameter (\citealt{Campitiello2022}) with highlighted in orange and blue the median profiles of mixed and disturbed targets, i.e. with a value of M below or above 0.4, respectively.}
    \label{fig:expected_rescaling}
\end{figure}
In the top panel of Fig.~\ref{fig:expected_rescaling}, we show the radio brightness profiles after the rescaling by the expected relations ($\gamma_M = 2.10$ and  $\gamma_z = 4.10$). It is worth noticing that by simply applying such rescaling (alongside the one for $R_{500}$) we reduce the intrinsic scatter by a factor $\sim 4$ (see Fig.\ref{fig:radio_prof}). These results may suggest that accounting properly for a mass and redshift dependence produces non-thermal component profiles with a common trend, resembling what has been observed for the thermal gas.\\
{In the bottom panel of Fig.~\ref{fig:expected_rescaling} we report the same rescaled profiles but color-coded by their dynamical state expressed through the M parameter.
In addition, since our sample covers rather well the CHEX-MATE distributions of disturbed and mixed targets as defined by \cite{Campitiello2022} (see Fig.~\ref{fig:comparison_plots}), i.e. the value of M is above or below 0.4 respectively, we can divide our sample according to such classification. Hence, we identified as mixed those objects having M<0.4 (8), and disturbed the ones with M>0.4 (8). 
We also computed the median radial profiles of the two subsamples, diving the radial range in regular bins and extracting the median surface brightness value for each of the two classes of objects. The median profiles are also displayed in the bottom panel Fig~\ref{fig:expected_rescaling}.}
It is clear how less disturbed clusters display a different and fainter halo profile, while, perturbed systems retain, on average, higher emission out to larger radii.
A difference in the integrated radio power between more relaxed and perturbed objects has been pointed out also by \cite{Cuciti21-2} and \cite{Cuciti2023}. 
Specifically, it was found that, regardless of cluster mass, the clusters that are more scattered in the $P_{\nu}-M$ relation are those that are more dynamically perturbed. This implies that cluster dynamics play a role in the scattering of the radio power-mass correlation.
In line with such results, we extend this scenario here, showing that the properties of merger-driven turbulence also have a key role in shaping the radio halo profiles.\\
The evidence of a dynamical dependence also suggests how, in future works and with wider samples, we can try to account for the cluster's dynamical status in the radial halo profile scaling model.
\subsection{Implications from the observed $R_H - M$ relation}\label{sec:no_RH-M}

We now try to improve our analyses further by revisiting and relaxing some of the assumptions made. To this aim, we exploit the observational evidence available for the considered LoTSS DR2 sample.\par
Inspired by self-similar scenario predictions, studies of the structure of the thermal ICM emitting in X-rays have found that there is a high degree of similarity on the radial distribution of the thermal properties when clusters are rescaled for radii at a given overdensity with respect to the critical density (e.g. \citealt{Lau2015,Ettori2020,Pratt2022} and references therein). 
In Sect.~\ref{sec:model}, we assumed that the halo size scales as the cluster mass and redshift, $R_H \propto M^{\beta_M} E_z^{\beta_z}$.
However, we do not know whether a relation of such kind holds also for the non-thermal component in clusters.
To the best of our knowledge, only \cite{Cassano2007} showed a superlinear relation of $R_H \propto R_{\Delta}^{2.63}$ and a strong relation between the radio halo size and cluster mass with a slope of $\sim 2.2$, indicating a clear no self-similarity of the non-thermal component.
In Table ~\ref{tab:spearman}, we present the Spearman rank among the main quantities considered in this work for the LoTSS DR2 clusters used to fit the $P_{\nu}-M$ relation.
In the left panel of Fig.~\ref{fig:Re-em-M}, instead, we plot the halo size ($R_H = 3 \times R_e$, as assumed in \citealt{Botteon2022}, that is the maximum integration radius for the radio halo profiles) versus the cluster mass. What is evident from both Table~\ref{tab:spearman} and Fig.~\ref{fig:Re-em-M} is that, at 150 MHz, there is no relation between the cluster mass and the halo size (see more in Sect.~\ref{sec:RH_discussion}).
However, given that the $P_{150MHz} - M_{500}$ relation exists, the parameter that must correlate with $M_{500}$ is the halo emissivity (Fig.~\ref{fig:Re-em-M}, right panel). 
{This can be explained in the framework of the turbulent re-acceleration, where, assuming that cluster mergers are responsible for the turbulence injection in the ICM, the synchrotron emissivity scales with the turbulent injection rate that depends on the cluster mass (e.g. \citealt{Cassano2005})}.\\
Hence, we relaxed the assumption made on $R_H \sim R_{\Delta}$ and generalised the model allowing for the radio halo size not to be correlated with the mass or redshift. We then consider the terms $\beta_M = \beta_z = 0$ in Eq.~\ref{eq:prop_relation}.
Considering this setup, Eq.~\ref{eq:rescaling_IR} can be rewritten as:
\begin{align}\label{eq:rescaling_IR-2}
    \log \left( \frac{I_{\nu}(r)}{(1+z)^{-(3+\alpha)}} \right ) =& ~ {\rm log}(I_0) + A \left( \frac{r}{r_{500}} \right) +
    \gamma_M \log \left( \frac{M_{500} } {10^{14} M_{\odot}} E_z\right)
\end{align}
where, in this case, $\gamma_M \equiv \alpha_M$. Thus, the profiles scale with the mass exactly as the integrated quantities. We report the results of the fit in Table~\ref{tab:Ir_fit}. This time the best-fit value of $\gamma_M = 2.88 \pm 0.33$ is well in agreement with the expected value of $2.76 \pm 0.35$.\par
On the other hand, as stated at the end of Sect.~\ref{sec:model}, we can infer which values of $\beta$ best accommodate our best-fit results (i.e., from Eq.~\ref{eq:rescaling_eqs}, $\beta_M = (\alpha_M-\gamma_M)/2$).
By combining the best-fit results in Tables~\ref{tab:P-M} and \ref{tab:Ir_fit}, we require values of $\beta_M$ between 0 and $-$0.4, with a typical error of 0.3, in a consistent manner with the independent evidence of a no-dependence upon the mass of the radio halo size.
\begin{figure*}
    \centering
    \includegraphics[scale=0.4]{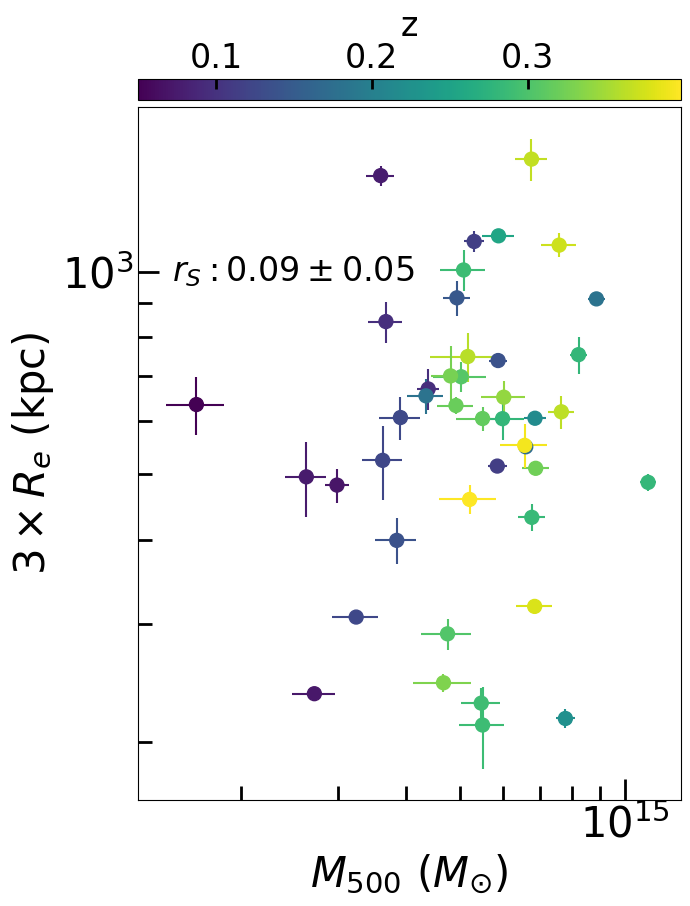}
    \includegraphics[scale=0.4]{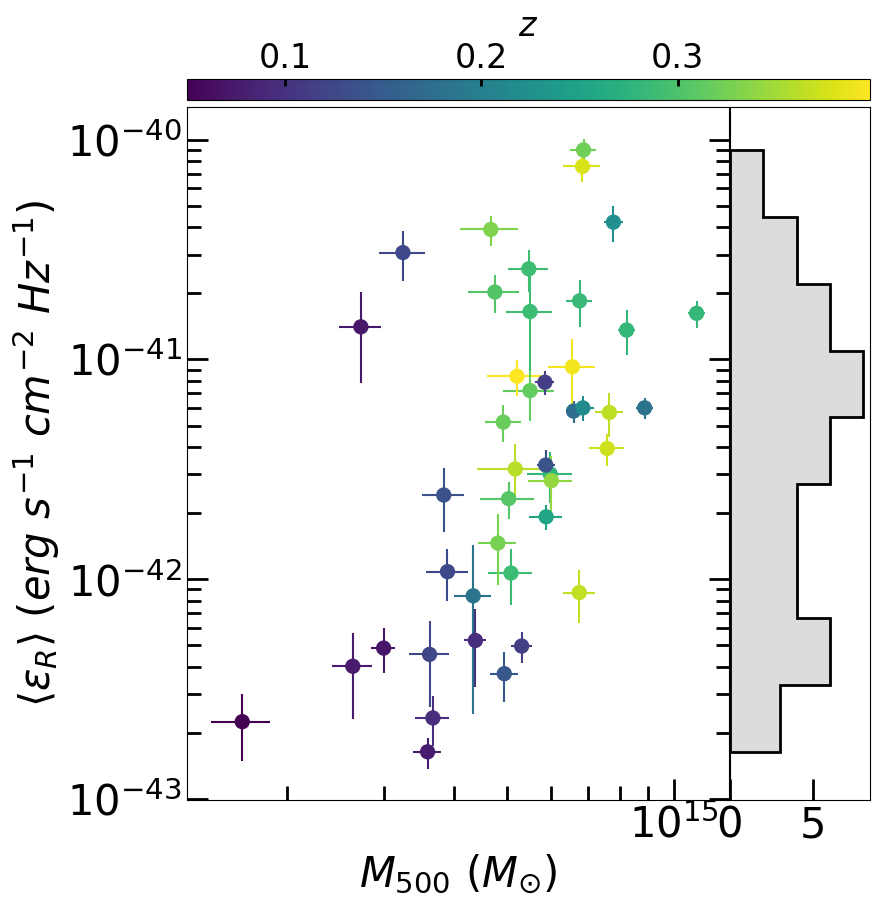}
    \caption{Correlations among three of the considered properties of the objects used to derive the $P_{150MHz} - M_{500}$ relations. Left: Mass vs radio halo size (expressed as $3 \times R_e$) and the Spearman rank. Right: Mass vs average radio emissivity, $\langle \epsilon_R \rangle$, considering the halo emission spherical and with a radius $3R_e$ reported in \cite{Botteon2022}.}
    \label{fig:Re-em-M}
\end{figure*}
\begin{table}
\centering
\caption{Spearman rank among the main quantities of the cluster sample studied by \cite{Cuciti2023} and also used here. Each cell is colored by the value within.}\label{tab:spearman}
\begin{tabular}{lcccccc}
\toprule
 & $z$ & $M_{500}$ & $R_{500}$ & $P_{150}$ & $I_0$ & $R_e$\\
\midrule
\midrule
$z$ & {\cellcolor[HTML]{FFFFFF}} \color[HTML]{000000}  & {\cellcolor[HTML]{8ED645}} \color[HTML]{000000} 0.55 & {\cellcolor[HTML]{A2DA37}} \color[HTML]{000000} 0.31 & {\cellcolor[HTML]{8BD646}} \color[HTML]{000000} 0.60 & {\cellcolor[HTML]{A0DA39}} \color[HTML]{000000} 0.33 & {\cellcolor[HTML]{C2DF23}} \color[HTML]{000000} -0.02 \\
$M_{500}$ & {\cellcolor[HTML]{8ED645}} \color[HTML]{000000} 0.55 & {\cellcolor[HTML]{FFFFFF}} \color[HTML]{000000}  & {\cellcolor[HTML]{77D153}} \color[HTML]{000000} 0.94 & {\cellcolor[HTML]{7FD34E}} \color[HTML]{000000} 0.80 & {\cellcolor[HTML]{8ED645}} \color[HTML]{000000} 0.56 & {\cellcolor[HTML]{B5DE2B}} \color[HTML]{000000} 0.09 \\
$R_{500}$ & {\cellcolor[HTML]{A2DA37}} \color[HTML]{000000} 0.31 & {\cellcolor[HTML]{77D153}} \color[HTML]{000000} 0.94 & {\cellcolor[HTML]{FFFFFF}} \color[HTML]{000000}  & {\cellcolor[HTML]{86D549}} \color[HTML]{000000} 0.70 & {\cellcolor[HTML]{93D741}} \color[HTML]{000000} 0.49 & {\cellcolor[HTML]{ADDC30}} \color[HTML]{000000} 0.17 \\
$P_{150}$ & {\cellcolor[HTML]{8BD646}} \color[HTML]{000000} 0.60 & {\cellcolor[HTML]{7FD34E}} \color[HTML]{000000} 0.80 & {\cellcolor[HTML]{86D549}} \color[HTML]{000000} 0.70 & {\cellcolor[HTML]{FFFFFF}} \color[HTML]{000000}  & {\cellcolor[HTML]{8BD646}} \color[HTML]{000000} 0.59 & {\cellcolor[HTML]{A2DA37}} \color[HTML]{000000} 0.28 \\
$I_0$ & {\cellcolor[HTML]{A0DA39}} \color[HTML]{000000} 0.33 & {\cellcolor[HTML]{8ED645}} \color[HTML]{000000} 0.56 & {\cellcolor[HTML]{93D741}} \color[HTML]{000000} 0.49 & {\cellcolor[HTML]{8BD646}} \color[HTML]{000000} 0.59 & {\cellcolor[HTML]{FFFFFF}} \color[HTML]{000000}  & {\cellcolor[HTML]{FDE725}} \color[HTML]{000000} -0.45 \\
$R_e$ & {\cellcolor[HTML]{C2DF23}} \color[HTML]{000000} -0.02 & {\cellcolor[HTML]{B5DE2B}} \color[HTML]{000000} 0.09 & {\cellcolor[HTML]{ADDC30}} \color[HTML]{000000} 0.17 & {\cellcolor[HTML]{A2DA37}} \color[HTML]{000000} 0.28 & {\cellcolor[HTML]{FDE725}} \color[HTML]{000000} -0.45 & {\cellcolor[HTML]{FFFFFF}} \color[HTML]{000000}  \\

\bottomrule
\end{tabular}
\end{table}

\section{Discussion}\label{sec:discussion}


%
%
 %

\subsection{On the $R_H-M$ relation}\label{sec:RH_discussion}

In Sect.~\ref{sec:no_RH-M}, we pointed out how, at LOFAR frequencies, there is no clear relation between the mass and the halo size. This, on one side, is in contrast with what was previously observed in higher-frequency studies (\citealt{Cassano2007,Cuciti21-2}), but on the other side might support the idea that, at such low frequencies, we are performing a wider census of the radio halo population that was previously inaccessible, indicating that such a relation may not hold on the whole population of radio halos.\\
However, before moving to the implications of our findings, we have to investigate and quantify possible biases in our analysis. 
\cite{Botteon2022} have already shown how, in low-surface brightness halos, the result of the exponential fit can be biased, leading the authors to exclude 10 clusters from all the subsequent analyses on LoTSS DR2 sample (labelling them as RH$^*$ and cRH$^*$).
Here, we investigate in greater detail the best-fit parameters of LoTSS DR2 targets searching for systematic effects that may influence the trend observed in the $R_H-M$ relation.\par
In Figure~\ref{fig:comparison_plots} we present the relation between
the $I_0/3\sigma_{RMS}$ ratio and $R_e$.
The black points in Figure~\ref{fig:comparison_plots} are the 10 excluded objects by the LoTSS DR2, while all the other points are the 43 LoTSS DR2 clusters, below $z=0.4$, considered in this work.
From this plot, we see that there is a cloud of points (magenta) that display large values of $R_e$ ($>150$ kpc) and low values (<4) of the $I_0/3\sigma_{RMS}$ ratio. It is interesting to see how these points are located close to the 10 excluded objects from the LoTSS DR2 (black points). This suggests that for the clusters in this range, the estimated size of the halo could have been influenced (particularly overestimated) by their low radio brightness.
In such a case, it would imply that, despite these targets having a sufficient radio signal to provide a good fit in terms of flux estimate (as reported in Appendix C in \citealt{Botteon2022}), the halo emission was not bright enough to provide an unbiased fit. 
\begin{figure}[h!]
    \centering
    \includegraphics[scale=0.55]{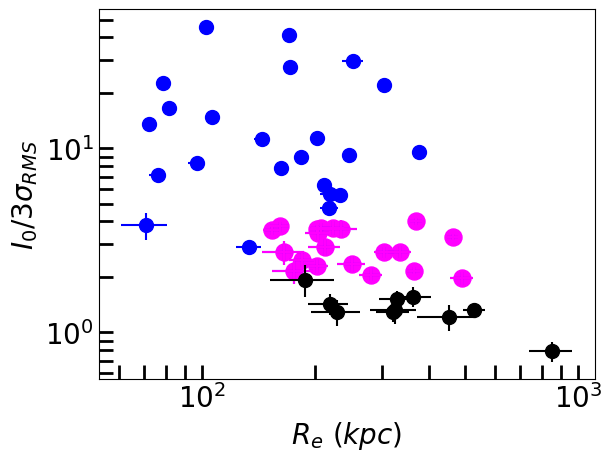}
    \caption{Comparison between $I_0/3\sigma_{RMS} - R_e$  for all LoTSS DR2 objects considered here, highlighting those targets (magenta) that lie in a similar range of parameters as the ten LoTSS DR2 objects (black) that had a non-reliable flux measurement (\cite{Botteon2022}).}
    \label{fig:comparison_plots}
\end{figure}
However, we stress that from Fig.~\ref{fig:comparison_plots} the main outcome is that there is no clear separation for the presence of objects with a biased exponential fit, but rather that there is a smooth transition from higher to lower signal-to-noise ratio (S/N) observations. Therefore, it is difficult to identify definitive criteria for selecting or rejecting an object (and the associated fit results). 
Instead, a careful investigation should be done to assess the reliability of different measurements and to determine their impact on the subsequent analyses and results, such as $R_H-M$ relation.\par
%
We can now highlight in the $R_H-M$ plane those targets for which we performed the aforementioned detailed analyses (top panel of Fig.~\ref{fig:RH-M-color}) and for which $R_e$ could have been overestimated.
\begin{figure}
    \centering
    \includegraphics[scale=0.5]{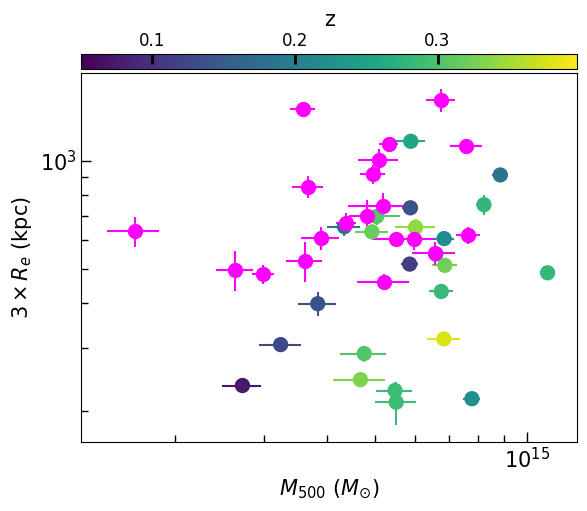}
    \includegraphics[scale=0.5]{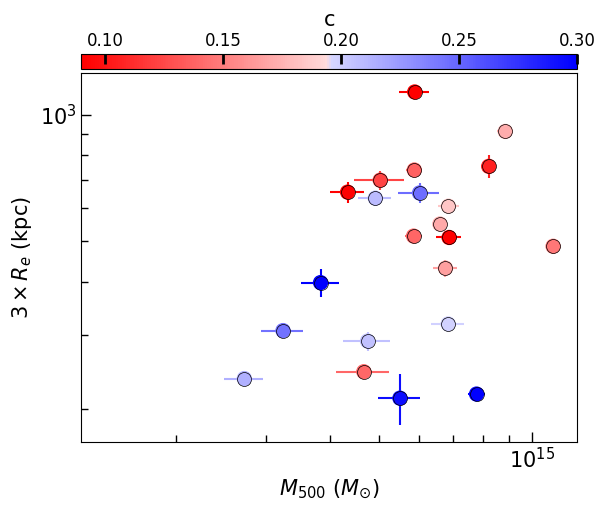}
    \caption{$3 R_e - M_{500}$ relation. Top: Same plot as left panel of Figure~\ref{fig:Re-em-M} but highlighting in magenta low S/N targets selected in Figure~\ref{fig:comparison_plots}. Bottom: Removing the magenta objects and color-coding the remaining ones according to $c$ from \cite{Zhang2023}.}
    \label{fig:RH-M-color}
\end{figure}
%
It is interesting to see that these targets have been selected by analysing parameters that are not related to the mass.
If removed from the $R_H-M$ plane, we find that the top left region of the plot is not populated anymore, suggesting the presence of an upper envelope of the observed $R_H$ at a given mass (Fig.~\ref{fig:RH-M-color}, bottom panel).
This can be explained by the fact that thanks to LOFAR, we can also detect the emission caused by minor merger events.
These events can occur both in low- and high-mass clusters, producing steep spectrum radio halos that will appear smaller in size (e.g. \citealt{Cassano2010}).
On the contrary, major merging events occur mainly in massive systems producing brighter and larger halos. As a consequence of that, we would have the formation of an upper envelope in the $R_H - M$ plane, as observed in Fig.~\ref{fig:RH-M-color}. \par
We can now exploit, when available, the dynamical parameter estimates for LoTSS DR2 objects to see any correlation in the $R_H-M$ plane. 
In particular, we make use of the results by \cite{Zhang2023}, who estimated such parameters within 100 and 500 kpc, differently to what \cite{Campitiello2022} did for the CHEX-MATE sample adopting 0.15 and 1 $R_{500}$.
Interestingly, color-coding the remaining targets by their $c$ parameters\footnote{We now use $c$ as a morphological indicator since, for the clusters in the LoTSS DR2 sample, only $c$ and $w$ were computed and not the full range of morphological parameters required to derive M ($c$, $w$, $P_{20}$ and $P_{30}$).} (Fig.\ref{fig:RH-M-color}, bottom panel) shows that only objects with a less concentrated core (i.e. higher dynamical disturbance) can have large halos and mainly at high masses. 
On the contrary, less massive and more relaxed targets display smaller halos. 
This latter finding using dynamical indicators further supports the view of the halo size being determined by both the cluster mass and the strength of the merger event in the system.
\subsection{Comparison with literature results}

We now discuss how our low-frequency results on the $R_H-M_{500}$ relation compare with previous findings.
In Figure~\ref{fig:fig:RH-M_C23-C21} (left panel), we compare the $3 R_e - M_{500}$ relation found here for the LoTSS DR2 targets with the one derived by using the results of \cite{Cuciti21-1} who fitted an exponential profile to a sample of halos observed at higher frequency\footnote{Note that the halo profile fit has not been performed with the same method as the one for LoTSS DR2 halos, which used Halo-FDCA}. 
We also compare, for the common targets of the two samples, the estimates for $R_e$ at different frequencies (right panel of Fig.~\ref{fig:fig:RH-M_C23-C21}). 
{ The cluster ZwCl0634.1+4750 it is not part of the LoTSS DR2 but was studied by \cite{Cuciti22} at 144 MHz obtaining $R_e \sim 150$kpc. Similarly, for A1758 we use the estimate provided by \cite{Balboni2024} as, due to a careful treatment of halo substructures, provides a more robust value of $R_e$ than the one reported in \cite{Botteon2022}.
} 
\begin{figure*}
    \centering
    \includegraphics[scale=0.5]{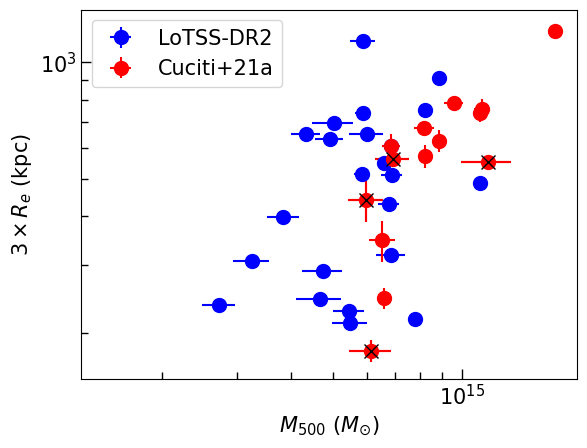}
    \includegraphics[scale=0.5]{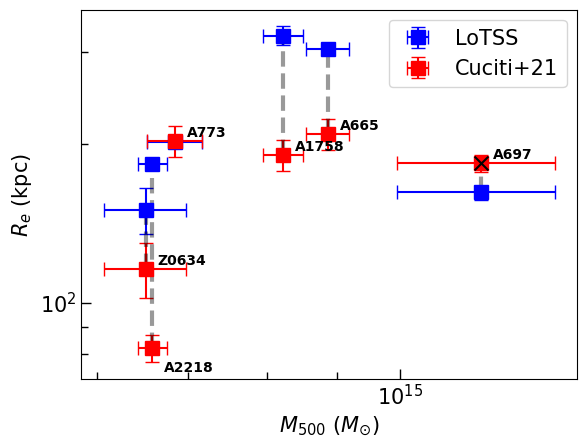}
    \caption{Comparison between the results of the LoTSS DR2 and from \cite{Cuciti21-1}. Left: The $3R_e - M_{500}$ relation (without the low-brightness points identified in this world). Right: Comparison of the estimated $R_e$ at low and high frequencies for the common objects. Points marked with a black 'X' are the ones for which GMRT data, instead of VLA, have been used. For the case of A773, only the red mark is visible as the two measures overlap.}
    \label{fig:fig:RH-M_C23-C21}
\end{figure*}\par
The most evident difference is that using VLA and GMRT observations, \cite{Cuciti21-1} recovered a correlation among the cluster mass and $R_e$ with a positive trend. 
Also, such a comparison clearly shows the differences between the two halo samples. 
First, the low-frequency observations allow us to explore the radio emission in low-mass systems, which are hardly detected in the GHz range.
Then, for a fixed mass, the extension of the halos is typically higher at lower frequencies. This suggests a broader distribution of the CRe at lower energies, making the halos appear more extended.
This is more evident in the right plot, where almost all clusters display more extended radio emission at lower frequencies\footnote{We note that for A697 the observed opposite trend of the halo size could be due to a poor estimation of $R_e$ by \cite{Cuciti21-1} caused by the presence of substructures within the halo ($\chi^2\sim 8$). The inversion between the low and high frequencies $R_e$ is no longer valid when it is observed with LOFAR at lower resolution, showing a wider radio emitting volume with $R_e \gtrsim 400$~kpc (\citealt{Cuciti22}).}. 
This could be interpreted as due to the presence of a radial spectral steepening in radio halos, which is naturally expected in the framework of the turbulence re-acceleration scenario (see e.g. \citealt{Brunetti14} and references therein). In fact, it is due to the re-acceleration of particles in a magnetic field with a decreasing radial profile, with the frequency at which the synchrotron steepens is $\nu_s\propto B$. 
This is observed in the case of few single radio halos (e.g. in the Coma cluster, \citealt{Bonafede2022}, MACSJ0717.5+3745; \citealt{Kamlesh-M0717}) however this is the first time this is derived for an ensemble of targets.\par
%
%
Finally, we have tested whether the objects with low S/N and a biased estimate of their size could impact our previous analyses. An incorrect estimate of $R_e$ would lead to an inaccurate estimate of the power since $P_{150MHz}$ is computed starting from model results. We have calculated the total radio power starting from the flux within the $2 \sigma_{RMS}$ halo region and derived the associated radio power. 
We re-computed the fit of the $P_{150MHz}-M_{500}$ relation made in Sect.~\ref{sec:m_only} finding no differences in the results ($\alpha_M = 3.07 \pm 0.38$) and we can conclude that our results are robust with respect to this systematic error.
\subsection{The role of the magnetic field in the $P-M$ relation}\label{sec:P-MB}
A power-law relation between the cluster total mass and the radio power associated with a giant halo has been confirmed and investigated in detail in recent works \citep[see e.g.][]{Cuciti21-2}. 
However, if the radio power associated with radio halos is interpreted as mainly induced by the magneto-hydrodynamical turbulence produced by mergers, the $P_{\nu}-M$ relation might include a dependence on the magnetic fields acting on the halo, and its form can become more complicated.\par
{ \cite{Cassano2005} have shown that radio halo can originate from merger-induced turbulence, where Kelvin-Helmholtz instabilities of the infalling substructures produce eddies in the ICM. They found that if a fraction of the turbulent energy is in the form of magnetosonic waves, the latter can re-accelerate relativistic electrons in the ICM, triggering synchrotron emission, both considering an injection spectrum over a broad range of scales or at a specific scale where the cascade is originated.
Following this modelling, \cite{Cassano06} have rewritten the radio power as: }
\begin{equation}\label{eq:P-MB}
    P_{\nu} \propto M^{2-\Gamma} \frac{B^2 \, n_{e, r}}{\left [ B^2 + B_{CMB}^2 \right ]^2},
\end{equation}
where $M$ is the cluster mass, $\Gamma$ is the dependence of the cluster temperature on the mass, $T \propto M^{\Gamma}$ ($\Gamma = 2/3$ in the self-similar scenario), $B = B_{M'} (M/M')^b$ is the magnetic field strength that scales with the object's mass $M$ as stated in \cite{Dolag2002,Dolag2004} who found $B \propto M^{1.33}$ for $\Gamma = 2/3$, $B_{CMB} \approx 3.25 (1+z)^2 \, \mu G$ is the equivalent magnetic field strength of the cosmic microwave background (CMB) and $n_{e, r}$ is the number density of relativistic electrons in the volume of the radio halos. 
In the two extreme regimes of $B_M \gg B_{CMB}$, and reverse, $P_{\nu} \approx M^{2-\Gamma} / \left[ M^{2b} +2b (1+z)^4 \right]$ and $P_{\nu} \approx M^{2-\Gamma +2b} / (1+z)^8$, respectively.\\
We model the radio power of the objects in the LoTSS DR2 sample at $z<0.4$ using of Eq.~\ref{eq:P-MB} and assuming $\Gamma=2/3$:
\begin{align}
    \log (P_{150MHz}) = ~ &N + \left( \frac{4}{3} - 2b \right) \log \frac{M}{M'} + \nonumber \\
   & -2 \log \left[ 1 + \left( \frac{B_{cmb,0}}{B_{M'}} \right)^2 \frac{(1+z)^4}{(M/M')^{2b}} \right],
   \label{eq:P-MB_log}
\end{align}
where $M' = 5.3 \times 10^{14} M_{\odot}$ is the median value of the distribution in mass in the sample and $B_{M'}$ is the average magnetic field at mass $M'$.
We then fit Eq.~\ref{eq:P-MB_log} to constrain the unknown parameters $B_{M'}$ and $b$. To do so, we used the python package \textsc{PyMC} v. 5.6.1 (\citealt{pymc16,pymc_v5.6.1,pymc}) which allows us to construct Bayesian models and fit them through Markov chain Monte Carlo methods, accounting for variables errors and intrinsic scatter (see Appendix~\ref{sec:appendix_fit} for more details). We obtain best-fit values, $B_{M'} = 2.52 \pm 0.61$ and $b=2.05 \pm 0.36$, and with a large intrinsic scatter of $0.95 \pm 0.12$.
We stress that these values refer to the radio halo emitting region, which, given that $P_{150MHz}$ is computed down to $3R_e$ (\citealt{Botteon2022}), is typically enclosed within $0.9-1 \rm ~ Mpc^3$.
The value of $b$ is higher than the reference value of 1.33 obtained by \cite{Dolag2002,Dolag2004} using cosmological simulations and also slightly larger than the constraints inferred by \cite{Cassano06}.
We can also test the case where $b$ is fixed to 1.33, finding a best-fit value for the magnetic field of $B_{M'}=2.11 \pm 0.88$ (see Figure~\ref{fig:alphaM_Bb+fit}).\par
Following \cite{Cassano06}, we can compare the expected value of the $P_{\nu} - M$ slope ($\alpha_M$) from the turbulent re-acceleration model with the one derived from observations. In this way, we can further determine the range of values for $B_{M'}$ and $b$ that are allowed by the observed $\alpha_M$.
The expected $P_{\nu} - M$ slope can be calculated as:
\begin{equation}
    \alpha_M = \frac{{\rm log}(P_1/P_2)}{{\rm log}(M_1/M_2)}
    \label{eq:P-M_slope}
\end{equation}
where for $M_1$ and $M_2$ we used the minimum and maximum mass value of the LoTSS DR2 clusters exploited here ($1.66 ~{\rm and}~ 10.10 \times 10^{14} M_{\odot}$), $P_1$ and $P_2$ are computed using Eq.~\ref{eq:P-MB} and assuming the mean redshift of the considered sample $\langle z \rangle = 0.23$. 
\begin{figure}
    \centering
    \includegraphics[scale=0.6]{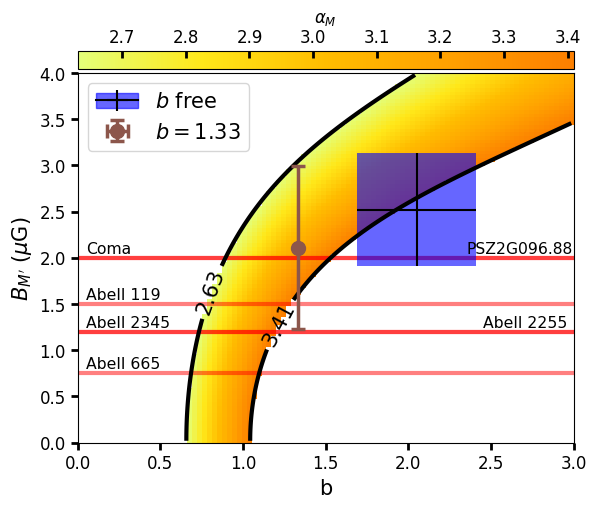}
    \caption{Expected $\alpha_M$ values as a function of $B_{M'}$ (y-axis) and $b$ (x-axis). In particular, the $\{B_{M'}, b\}$ range shown (in orange) is the one allowed by the $\alpha_M$ slope obtained from our analysis made in Sect.~\ref{sec:m_only} (in orange). The blue box and brown point are the best-fit values from the fit of Eq.~\ref{eq:P-MB_log}.
    Red lines represent literature results of estimates of the average cluster magnetic field from polarisation studies (\citealt{Murgia2004, Govoni2006, Guidetti2008, Bonafede2010,Vacca2010, Vacca2012, Govoni2017,Stuardi2021,DeRubeis2024}).}
    \label{fig:alphaM_Bb+fit}
\end{figure}
In Figure~\ref{fig:alphaM_Bb+fit}, we report the derived $\alpha_M$ obtained from Eq.~\ref{eq:P-M_slope} for the different combination of $B_M$ and $b$, which lie in the $\{B_{M'}, b\}$ parameters space allowed by our fit of the $P_{150MHz} - M_{500}$ relation (see Sect.\ref{sec:m_only}).
Additionally, since we fitted Eq.~\ref{eq:P-MB_log} to the data, we also report the recovered best-fit values for $B_{M'}$ and $b$. 
The blue box indicates the results when leaving all the parameters free to vary and the brown point is for the case with a fixed $b=1.33$.
We see that both results are consistent with the ones we derived by fitting the $P_{\nu} - M_{500}$ relation.
However, the fit obtained with a fixed $b$ agrees better with the $\{B_{M'}, b\}$ range allowed by $\alpha_M$ (yellow-orange region) than the one with $b$ free to vary, whose consistency appears more marginal.
We also note that the intrinsic scatter of the former is higher $\sim 1.04$ than that of the latter $\sim 0.95$.\par
We compare in Fig.\ref{fig:alphaM_Bb+fit} our estimates of $B_{M'}$ with the literature results and analyses reported in \cite{Stuardi2021} and \cite{DeRubeis2024} about clusters' magnetic field estimates within a volume of 1 Mpc$^3$. 
We focus only on those objects classified as merging systems. 
Those estimates have been obtained through polarisation studies of compact sources embedded within the cluster environment or background sources at higher redshift and subsequent comparisons with simulations (the only exception is the work of \citealt{DeRubeis2024}, which used only the polarised emission from radio relics).
It is important to note that our values of $B_{M'}$ are referred to a cluster with a mass of $5.3 \times 10^{14} M_{\odot}$, while such literature results span a range of masses, $\sim 3.4 - 8.9 \times 10^{14} M_{\odot}$, and that include clusters with and without radio halos (at the frequency used to determine the cluster magnetic field).
Bearing this in mind, we see that, although they lie in the same $\sim \mu$G range, our best-fit values are higher than the ones from the literature by a factor of 1.5-2, with only Coma and PSZ2G096.88 that are consistent with both our fits.
We note that of the two estimates obtained here, the fit with fixed $b=1.33$, which predicts a lower value of $B_{M'}$, agrees better with past results.
A possible explanation for the discrepancy between the two approaches could be the different cluster volumes in which the average $B$ is estimated. In particular, literature results all refer to a volume of 1 Mpc$^3$, while in our sample several clusters have a smaller extension and the magnetic field is expected to decline with the radius (\citealt{Bonafede2010}). 

\section{Summary and conclusions}\label{sec:conclusion}


In this work, we analysed the global and spatially resolved non-thermal properties of a selected sample of radio halos in terms of their dependence with mass and redshift, inspired by the analogy of what has been achieved for the ICM thermal components. With such analyses, we unveiled the complex scenario brought up by low-frequency observations and also put constraints
on the role of the magnetic field.\par
We initially derived a simple model that links the total radio power to the halo surface brightness profile, relating the dependence with mass and redshift of the global quantity to the one of the spatially resolved profile (Eq.~\ref{eq:P-Mz} and Eq.~\ref{eq:rescaling_IR}).
We used this model to obtain the expected scaling (in $M$ and $z$) of the radio profiles starting from the observed dependencies of the total radio power (through Eq.~\ref{eq:rescaling_eqs}).
We then exploited the available radio halo profiles to recover the best-fit scaling parameters from the data and compare them with the ones predicted by our model based on the radio power-mass relation.
With these tools, we explored different possibilities for the mass and redshift scaling relations.
We started by considering only the mass dependence in our relations, finding that the scatter of the profiles is minimised ($\sigma_{int} \sim 0.22$). However, the best-fit parameters obtained from the observed profiles show a disagreement (at $2\sigma$ level) with the model expectations. 
We then accounted also for the redshift dependence, finding an improvement in the results, but still with a discrepancy (at $1.4 \sigma$) with model predictions. 
Despite that, when rescaling the halo profiles for the expected dependencies of the model, the profiles' scatter was reduced by a factor of 4.
We also reported an evident residual dependence of the scaled halo profile on the cluster dynamical status (Fig.\ref{fig:expected_rescaling}).\par
However, since no $R_H-M$ correlation is found for the LoTSS DR2 clusters, we then relaxed the constraint on the halo size mass dependence and modified the considered model (Eq.\ref{eq:rescaling_IR-2}).
In this way, we recovered a good agreement between the mass profile scaling predicted by the model ($\sim 2.76$) and the one recovered from the best-fit parameters ($\sim 2.88$), whilst still significantly reducing the profiles' intrinsic scatter ($\sigma_{int} \sim 0.26$).\par
In Sect.~\ref{sec:RH_discussion}, we analysed and discussed the $R_H-M$ relation, finding that for some low S/N objects, the halo size could have been overestimated. 
When removed, in the $R_H-M$ relation it is possible to identify an upper envelope for the size of the radio halo that does not allow for large radio halos at low cluster masses.
We interpreted this evidence as an indication of the dependence of the halo size on the energetics of the merger, with only clusters that experienced minor merger events being found at low masses, while more massive clusters can host both large and small halos depending on the type of the merger event, as expected in the turbulent re-acceleration scenario (e.g. \citealt{Cassano2010}) .\par
Finally, we tried to assess the role of the magnetic field in the observed $P_{\nu}-M$ relation by exploiting the work from \cite{Cassano06} and, thanks to our larger sample, expanding their analysis. 
Using the model described in their work we estimated the scaling with the mass of the volume-averaged cluster magnetic field ($B \propto M^b$), finding a value of $b=2.05 \pm 0.36$ and an average magnetic field $B_{M'} = 2.52 \pm 0.61$ for $M'=5.3 \times 10^{14} M_{\odot}$. 
When compared with the results discussed in Sect.~\ref{sec:m_only}, those estimates are in agreement with the $\{B_{M'}, b\}$ parameter space allowed by the observed $P_{150}-M_{500}$ correlation slope.
In addition, our results of $B_{M'}$ lie in the $1-2 ~\mu$G range, in agreement with literature results. However, previous estimates of the average cluster magnetic field are typically lower than the ones found here. We suggest that this can be due to different effects, ranging from sample selection (e.g. mass, presence of a radio halo) to the differences in the considered cluster volume. 
We also highlight the capabilities of studies like the one presented here in constraining the cluster magnetic field through the scaling relations in halos. In fact, they can provide complementary and independent results to the ones obtained through polarisation cluster analyses, the more commonly used approach to constrain the cluster magnetic field.
\begin{acknowledgements}
We acknowledge the developers of the following Python packages which were used in this work: \textsc{ASTROPY} \citep{astropy:2013,astropy:2018,astropy:2022}, \textsc{MATPLOTLIB} \citep{matplotlib}, \textsc{SCIPY} \citep{SciPy-NMeth}, \textsc{NUMPY} \citep{numpy} and \textsc{CORNER}, (\citealt{cornerpy16}). 
LOFAR \citep{LOFAR2013} is the Low Frequency Array designed and constructed by ASTRON. It has observing, data processing, and data storage facilities in several countries, which are owned by various parties (each with their own funding sources) and are collectively operated by the ILT foundation under a joint scientific policy. The ILT resources have benefited from the following recent major funding sources: CNRS-INSU, Observatoire de Paris and Université d'Orléans, France; BMBF, MIWF-NRW, MPG, Germany; Science Foundation Ireland (SFI), Department of Business, Enterprise and Innovation (DBEI), Ireland; NWO, The Netherlands; The Science and Technology Facilities Council, UK; Ministry of Science and Higher Education, Poland; The Istituto Nazionale di Astrofisica (INAF), Italy. This research made use of the Dutch national e-infrastructure with the support of the SURF Cooperative (e-infra 180169) and the LOFAR e-infra group. The Jülich LOFAR Long Term Archive and the GermanLOFAR network are both coordinated and operated by the Jülich Supercomputing Centre (JSC), and computing resources on the supercomputer JUWELS at JSC were provided by the Gauss Centre for Supercomputinge.V. (grant CHTB00) through the John von Neumann Institute for Computing (NIC). This research made use of the University of Hertfordshirehigh-performance computing facility and the LOFAR-UK computing facility located at the University of Hertfordshire and supported by STFC [ST/P000096/1], and of the Italian LOFAR IT computing infrastructure supported and operated by INAF, and by the Physics Department of Turin university (under an agreement with Consorzio Interuniversitario per la Fisica Spaziale) at the C3S Supercomputing Centre, Italy.\\
S.E., F.G., M.R., I.B. acknowledge the financial contribution from the contracts
Prin-MUR 2022 supported by Next Generation EU (M4.C2.1.1, n.20227RNLY3 {\it The concordance cosmological model: stress-tests with galaxy clusters}),
ASI-INAF Athena 2019-27-HH.0, ``Attivit\`a di Studio per la comunit\`a scientifica di Astrofisica delle Alte Energie e Fisica Astroparticellare'' (Accordo Attuativo ASI-INAF n. 2017-14-H.0).
S.E. also acknowledges support from the European Union’s Horizon 2020 Programme under the AHEAD2020 project (grant agreement n. 871158).
M.G. acknowledges support from the ERC Consolidator Grant \textit{BlackHoleWeather} (101086804).
B.J.M. acknowledges support from STFC grant ST/Y002008/1.
E.R. acknowledges support from the Chandra Theory Program (TM4-25006X) awarded from the Chandra X-ray Center which is operated by SAO for and
on behalf of NASA under contract NAS8-03060.
E.P. acknowledges support from the French Agence Nationale de la Recherche (ANR), under grant ANR-22-CE31-0010.
J.S. was supported by NASA Astrophysics Data Analysis Program (ADAP) Grant 80NSSC21K1571
R.J.vW. acknowledges support from the ERC Starting Grant ClusterWeb 804208. 
R.C. acknowledges financial support from the INAF grant 2023
``Testing the origin of giant radio halos with joint LOFAR" (1.05.23.05.11).
\end{acknowledgements}

%
\bibliographystyle{aa} 
\bibliography{biblio} 
%

\begin{appendix} 

\section{Fit of the P-M-B relation}\label{sec:appendix_fit}

We fitted Eq.\ref{eq:P-MB_log} using the python package \textit{PyMC} \citep{pymc16, pymc_v5.6.1} which creates Bayesian models and fit them through Markov chain Monte Carlo methods. In particular, we constructed a model that accounts for x and y errors and includes the intrinsic scatter. We adopted uniform priors on both $B_{M'}$ and $b$ and considered both the mass and the total radio power normally distributed (e.g. \citealt{Sereno2016}). The best-fit parameters for the two considered cases (with and without fixing $b$ to 1.33) are reported in Table~\ref{tab:fit_par}).\\
\begin{table}[ht!]
    \centering
    \begin{tabular}{l l c}
        Parameter       & Prior     & Best fit \\
        \hline
        \hline
        $b$             & $U(0,10)$ & $2.05 \pm 0.36$ \\
        $B_{M'}$        & $U(0,10)$ & $2.53 \pm 0.61$ \\
        $N$             & $U(0,10)$ & $2.82 \pm 0.83$ \\
        $\sigma_{int}$  & $HC(0.5)$ & $0.95 \pm 0.13$ \\
        \hline
        $b$ (fixed)     & /         & 1.33 \\
        $B_{M'}$        & $U(0,10)$ & $2.11 \pm 0.88$ \\
        $N$             & $U(0,10)$ & $3.20 \pm 1.69$ \\
        $\sigma_{int}$  & $HC(0.5)$ & $1.04 \pm 0.13$ \\
    \end{tabular}
    \caption{List of the parameters used in the fit of Eq/\ref{eq:P-MB_log}, the priors used (uniform and half-Cauchy), and the best-fit value.}
    \label{tab:fit_par}
\end{table}
We also report in Figure~\ref{fig:appendix_ctrplot} the contour plots for the case with all the parameters free to vary. It is clear how the only degeneration is present between the normalisation $N$ and the average cluster magnetic field $B_{M'}$. This is somehow expected since $B_{M'}$ plays indeed the role of a normalisation in the $B(M) = B_{M'} (M/M')^b$ relation included in Eq.\ref{eq:P-MB_log}. Therefore, the two parameters appear anti-correlated.
\begin{figure}[h]
    \centering
    \includegraphics[width=\linewidth]{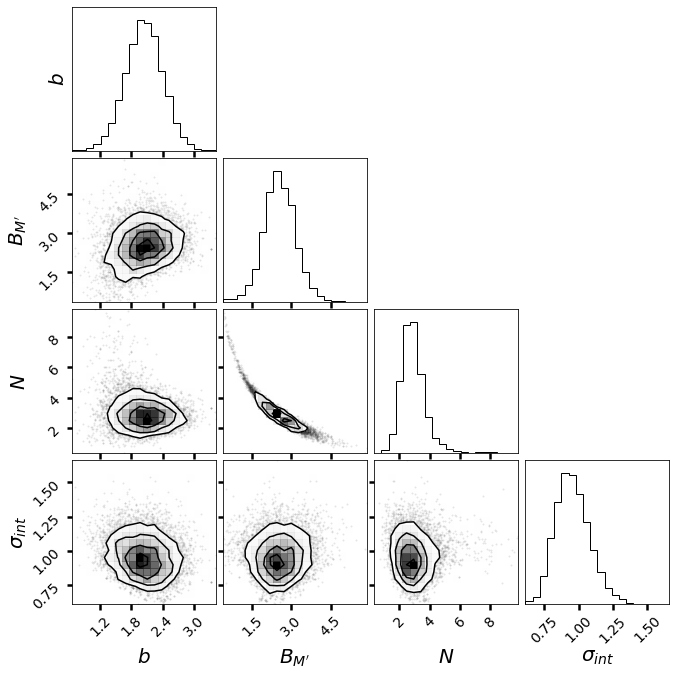}
    \caption{Contour plots of the fitted variables of Eq.\ref{eq:P-MB_log} with the median values of each distribution highlighted.}
    \label{fig:appendix_ctrplot}
\end{figure}

\section{General properties of the LoTSS DR2 halo clusters}\label{appendix:table_RH}

\begin{table*}[ht!]
    \centering
    \begin{tabular}{lcccccccccc}
    \toprule
    Name & $z$ & \thead{$M_{500}$ \\ $(10^{14} {M_{\odot}})$ } & \thead{ $P_{150 \rm MHz}$ \\ $\rm ( 10^{24} ~ W~Hz^{-1})$} & \thead{$I_0$ \\ ($\mu$Jy arcsec$^{-2}$)} & $R_e ~ {\rm (kpc)}$ & $c$ & $w~ ( 10^{-1})$ \\
    \midrule
    PSZ2G023.17+86.71 & 0.306 & $5.03 \pm 0.56$ & $9.76 \pm 1.18$ & $5.79 \pm 0.34$ & $232.8 \pm 11.7$ & $0.123 \pm 0.010$ & $0.217 \pm 0.026$ \\
    PSZ2G031.93+78.71 & 0.072 & $2.72 \pm 0.24$ & $2.28 \pm 1.01$ & $11.39 \pm 0.20$ & $78.7 \pm 1.1$ & $0.214 \pm 0.002$ & $0.283 \pm 0.002$ \\
    PSZ2G040.58+77.12 & 0.075 & $2.63 \pm 0.22$ & $0.60 \pm 0.11$ & $0.67 \pm 0.10$ & $165.3 \pm 21.0$ & $0.227 \pm 0.006$ & $0.061 \pm 0.007$ \\
    PSZ2G045.87+57.70 & 0.611 & $7.03 \pm 0.68$ & $21.1 \pm 5.03$ & $14.18 \pm 1.90$ & $89.0 \pm 9.6$ & $0.254 \pm 0.005$ & $0.218 \pm 0.007$ \\
    PSZ2G046.88+56.48 & 0.115 & $5.31 \pm 0.23$ & $8.36 \pm 0.98$ & $1.57 \pm 0.07$ & $369.9 \pm 13.6$ & $0.082 \pm 0.004$ & $0.234 \pm 0.018$ \\
    PSZ2G048.10+57.16 & 0.078 & $3.59 \pm 0.21$ & $5.43 \pm 0.68$ & $1.58 \pm 0.05$ & $463.4 \pm 15.6$ & $0.088 \pm 0.003$ & $0.591 \pm 0.076$ \\
    PSZ2G049.32+44.37 & 0.097 & $3.67 \pm 0.26$ & $1.73 \pm 0.26$ & $0.61 \pm 0.05$ & $281.2 \pm 19.7$ & $0.184 \pm 0.006$ & $0.109 \pm 0.011$ \\
    PSZ2G053.53+59.52 & 0.113 & $5.85 \pm 0.23$ & $13.20 \pm 1.56$ & $11.66 \pm 0.11$ & $171.5 \pm 1.4$ & $0.139 \pm 0.003$ & $0.132 \pm 0.035$ \\
    PSZ2G055.59+31.85 & 0.224 & $7.78 \pm 0.31$ & $5.29 \pm 0.86$ & $17.48 \pm 0.71$ & $72.3 \pm 2.3$ & $0.300 \pm 0.015$ & $0.056 \pm 0.038$ \\
    PSZ2G056.77+36.32 & 0.095 & $4.38 \pm 0.20$ & $1.95 \pm 0.63$ & $1.10 \pm 0.10$ & $223.3 \pm 15.5$ & $0.303 \pm 0.010$ & $0.039 \pm 0.018$ \\
    PSZ2G066.41+27.03 & 0.576 & $7.70 \pm 0.53$ & $172.00 \pm 19.50$ & $15.5 \pm 0.42$ & $374.5 \pm 11.0$ & $0.088 \pm 0.009$ & $0.275 \pm 0.228$ \\
    PSZ2G071.21+28.86 & 0.366 & $6.75 \pm 0.45$ & $34.10 \pm 5.44$ & $1.53 \pm 0.13$ & $490.3 \pm 35.6$ & $0.064 \pm 0.004$ & $0.129 \pm 0.021$ \\
    PSZ2G083.29-31.03 & 0.412 & $8.27 \pm 0.44$ & $36.70 \pm 4.03$ & $6.71 \pm 0.28$ & $226.5 \pm 7.8$ & $0.177 \pm 0.013$ & $0.297 \pm 0.107$ \\
    PSZ2G084.13-35.41 & 0.314 & $5.50 \pm 0.58$ & $19.70 \pm 4.69$ & $6.16 \pm 0.31$ & $201.8 \pm 8.4$ & $0.095 \pm 0.006$ & $0.379 \pm 0.021$ \\
    PSZ2G096.83+52.49 & 0.318 & $4.92 \pm 0.37$ & $16.20 \pm 2.80$ & $7.92 \pm 0.23$ & $211.0 \pm 5.9$ & $0.209 \pm 0.004$ & $0.087 \pm 0.009$ \\
    PSZ2G097.72+38.12 & 0.171 & $6.59 \pm 0.16$ & $11.90 \pm 1.27$ & $7.44 \pm 0.14$ & $183.2 \pm 2.9$ & $0.170 \pm 0.007$ & $0.321 \pm 0.079$ \\
    PSZ2G099.86+58.45 & 0.630 & $6.85 \pm 0.49$ & $38.70 \pm 9.21$ & $7.36 \pm 0.49$ & $163.2 \pm 8.6$ & $0.133 \pm 0.010$ & $0.215 \pm 0.056$ \\
    PSZ2G106.61+66.71 & 0.331 & $4.67 \pm 0.56$ & $7.07 \pm 0.90$ & $12.82 \pm 0.52$ & $81.7 \pm 2.5$ & $0.140 \pm 0.032$ & $0.508 \pm 0.080$ \\
    PSZ2G107.10+65.32 & 0.280 & $8.22 \pm 0.28$ & $71.60 \pm 9.60$ & $22.66 \pm 1.44$ & $251.1 \pm 16.0$ & $0.107 \pm 0.006$ & $0.861 \pm 0.012$ \\
    PSZ2G109.97+52.84 & 0.326 & $4.81 \pm 0.38$ & $6.18 \pm 0.98$ & $1.39 \pm 0.16$ & $233.5 \pm 25.0$ & $0.334 \pm 0.005$ & $0.082 \pm 0.009$ \\
    PSZ2G111.75+70.37 & 0.183 & $4.34 \pm 0.33$ & $2.90 \pm 1.99$ & $2.53 \pm 0.14$ & $218.1 \pm 12.9$ & $0.092 \pm 0.007$ & $0.571 \pm 0.032$ \\
    PSZ2G112.48+56.99 & 0.070 & $2.99 \pm 0.15$ & $0.67 \pm 0.10$ & $0.80 \pm 0.05$ & $160.7 \pm 9.3$ & $0.174 \pm 0.005$ & $0.046 \pm 0.01$ \\
    PSZ2G113.91-37.01 & 0.371 & $7.58 \pm 0.55$ & $63.8 \pm 7.09$ & $5.09 \pm 0.23$ & $365.3 \pm 15.0$ & $0.157 \pm 0.016$ & $0.460 \pm 0.025$ \\
    PSZ2G114.31+64.89 & 0.284 & $6.76 \pm 0.37$ & $18.3 \pm 3.64$ & $12.40 \pm 0.59$ & $143.9 \pm 6.6$ & $0.166 \pm 0.027$ & $0.128 \pm 0.019$ \\
    PSZ2G133.60+69.04 & 0.254 & $5.88 \pm 0.40$ & $34.3 \pm 4.28$ & $7.07 \pm 0.17$ & $377.1 \pm 5.2$ & $0.087 \pm 0.009$ & $0.380 \pm 0.035$ \\
    PSZ2G135.17+65.43 & 0.544 & $6.01 \pm 0.60$ & $54.6 \pm 9.84$ & $6.43 \pm 0.5$ & $232.7 \pm 12.1$ & $0.105 \pm 0.019$ & $0.472 \pm 0.076$ \\
    PSZ2G138.32-39.82 & 0.280 & $5.98 \pm 0.56$ & $8.22 \pm 1.27$ & $2.89 \pm 0.24$ & $201.7 \pm 14.3$ & $0.198 \pm 0.007$ & $0.132 \pm 0.013$ \\
    PSZ2G139.18+56.37 & 0.322 & $6.87 \pm 0.38$ & $147.0 \pm 18.40$ & $45.75 \pm 0.25$ & $170.2 \pm 0.8$ & $0.086 \pm 0.006$ & $0.470 \pm 0.085$ \\
    PSZ2G143.26+65.24 & 0.363 & $7.65 \pm 0.43$ & $16.8 \pm 2.53$ & $4.30 \pm 0.27$ & $206.5 \pm 11.7$ & $0.142 \pm 0.026$ & $0.246 \pm 0.015$ \\
    PSZ2G148.36+75.23 & 0.304 & $4.75 \pm 0.50$ & $6.07 \pm 0.70$ & $8.57 \pm 0.58$ & $96.6 \pm 5.1$ & $0.206 \pm 0.009$ & $0.527 \pm 0.024$ \\
    PSZ2G149.22+54.18 & 0.137 & $5.87 \pm 0.22$ & $16.4 \pm 2.69$ & $6.42 \pm 0.11$ & $246.1 \pm 3.6$ & $0.136 \pm 0.003$ & $0.037 \pm 0.008$ \\
    PSZ2G149.75+34.68 & 0.182 & $8.86 \pm 0.32$ & $56.2 \pm 6.04$ & $12.23 \pm 0.08$ & $303.9 \pm 1.6$ & $0.172 \pm 0.003$ & $0.613 \pm 0.037$ \\
    PSZ2G150.56+58.32 & 0.470 & $7.55 \pm 0.51$ & $56.5 \pm 10.70$ & $10.66 \pm 0.37$ & $315.3 \pm 11.2$ & $0.133 \pm 0.016$ & $0.316 \pm 0.177$ \\
    PSZ2G151.19+48.27 & 0.289 & $5.08 \pm 0.47$ & $13.4 \pm 2.57$ & $1.66 \pm 0.13$ & $335.5 \pm 23.7$ & $0.077 \pm 0.009$ & $0.241 \pm 0.118$ \\
    PSZ2G164.65+46.37 & 0.342 & $6.01 \pm 0.55$ & $9.56 \pm 2.26$ & $5.87 \pm 0.33$ & $217.1 \pm 12.0$ & $0.246 \pm 0.010$ & $0.605 \pm 0.021$ \\
    PSZ2G165.06+54.13 & 0.144 & $4.94 \pm 0.28$ & $3.51 \pm 0.62$ & $0.88 \pm 0.06$ & $305.0 \pm 18.4$ & $0.188 \pm 0.005$ & $0.177 \pm 0.015$ \\
    PSZ2G166.09+43.38 & 0.217 & $6.85 \pm 0.32$ & $16.5 \pm 1.87$ & $7.19 \pm 0.16$ & $202.0 \pm 3.9$ & $0.184 \pm 0.007$ & $0.183 \pm 0.057$ \\
    PSZ2G172.63+35.15 & 0.127 & $3.90 \pm 0.34$ & $2.99 \pm 0.41$ & $1.79 \pm 0.16$ & $202.5 \pm 15.1$ & $0.184 \pm 0.009$ & $0.201 \pm 0.021$ \\
    PSZ2G176.27+37.54 & 0.567 & $6.14 \pm 0.86$ & $9.62 \pm 3.16$ & $2.96 \pm 1.05$ & $139.6 \pm 40.7$ & $0.243 \pm 0.017$ & $0.190 \pm 0.039$ \\
    PSZ2G179.09+60.12 & 0.137 & $3.84 \pm 0.33$ & $1.89 \pm 0.42$ & $2.55 \pm 0.22$ & $133.0 \pm 10.4$ & $0.515 \pm 0.006$ & $0.065 \pm 0.022$ \\
    PSZ2G183.90+42.99 & 0.561 & $6.95 \pm 0.74$ & $75.5 \pm 7.81$ & $16.35 \pm 0.39$ & $167.6 \pm 3.1$ & $0.156 \pm 0.005$ & $0.182 \pm 0.011$ \\
    PSZ2G184.68+28.91 & 0.288 & $5.50 \pm 0.52$ & $1.94 \pm 0.40$ & $5.40 \pm 0.93$ & $70.7 \pm 9.8$ & $0.293 \pm 0.015$ & $0.079 \pm 0.03$ \\
    PSZ2G186.37+37.26 & 0.282 & $11.00 \pm 0.37$ & $23.0 \pm 2.44$ & $12.40 \pm 0.43$ & $162.2 \pm 4.8$ & $0.147 \pm 0.008$ & $0.099 \pm 0.053$ \\
    PSZ2G186.99+38.65 & 0.378 & $6.84 \pm 0.52$ & $30.0 \pm 3.96$ & $27.68 \pm 0.90$ & $106.1 \pm 2.8$ & $0.199 \pm 0.008$ & $0.385 \pm 0.021$ \\
    PSZ2G189.31+59.24 & 0.126 & $3.24 \pm 0.31$ & $10.9 \pm 2.78$ & $25.8 \pm 0.39$ & $102.3 \pm 1.2$ & $0.245 \pm 0.004$ & $0.476 \pm 0.008$ \\
    PSZ2G190.61+66.46 & 0.488 & $5.55 \pm 0.65$ & $36.3 \pm 4.44$ & $3.22 \pm 0.23$ & $290.5 \pm 15.0$ & $0.105 \pm 0.016$ & $0.287 \pm 0.058$ \\
    PSZ2G192.18+56.12 & 0.124 & $3.62 \pm 0.30$ & $0.81 \pm 0.15$ & $0.66 \pm 0.10$ & $174.9 \pm 22.1$ & $0.172 \pm 0.007$ & $0.170 \pm 0.112$ \\
    PSZ2G205.90+73.76 & 0.447 & $7.39 \pm 0.55$ & $30.1 \pm 5.15$ & $34.30 \pm 1.53$ & $103.3 \pm 4.7$ & $0.212 \pm 0.018$ & $0.135 \pm 0.032$ \\                
    \bottomrule
    \end{tabular}
    \caption{Main properties of the 43 radio halo clusters at z < 0.4 used to constrain the $P_{\nu}-M$ relation taken from \cite{Cuciti2023} (and with X-ray morphological parameters in \cite{Zhang2023}), the sample listes as the first entry in Table~\ref{tab:samples}.}
    \label{tab:lotssdr2_sample}
\end{table*}

\end{appendix}
\end{document}